\pdfoutput=1
\documentclass[twocolumn,twocolappendix, numberedappendix]{aastex6}
\usepackage{natbib, threeparttablex, amsmath, gensymb, longtable, hyperref}

\newcommand\vlsr{v_{\rm LSR}}
\newcommand\Msun{\; {\rm M}_{\odot}}
\newcommand\kms{\; {\rm km}\;{\rm s}^{-1}}
\newcommand\Msunyr{\; {\rm M}_{\odot}\; {\rm yr}^{-1}}

\usepackage{array}
\newcolumntype{X}[1]{>{\centering\arraybackslash}m{#1}}

\begin{document}
\shorttitle{Gas Accretion at the Disk-halo Interface of M33}
\shortauthors{Zheng et al.}
\title{HST/COS Observations of Ionized Gas Accretion at the Disk-Halo Interface of M33}
\author{Y. Zheng$^1$, J. E. G. Peek$^2$, J. K. Werk$^3$, M. E. Putman$^1$}
\affil{$^1$ Department of Astronomy, Columbia University, New York, NY 10027, USA; yzheng@astro.columbia.edu \\
       $^2$ Space Telescope Science Institute, 3700 San Martin Dr, Baltimore, MD 21218, USA \\
       $^3$ Department of Astronomy, University of Washington, Seattle, WA 98195-1580, USA}
\begin{abstract}

We report the detection of accreting ionized gas at the disk-halo interface of the nearby galaxy M33. We analyze {\it HST}/COS absorption-line spectra of seven ultraviolet-bright stars evenly distributed across the disk of M33. We find \ion{Si}{4} absorption components consistently redshifted relative to the bulk M33's ISM absorption along all the sightlines. The \ion{Si}{4} detection indicates an enriched, disk-wide, ionized gas inflow toward the disk. This inflow is most likely multi-phase as the redshifted components can also be observed in ions with lower ionization states (e.g., \ion{S}{2}, \ion{P}{2}, \ion{Fe}{2}, \ion{Si}{2}). Kinematic modeling of the inflow is consistent with an accreting layer at the disk-halo interface of M33, which has an accretion velocity of 110$^{+15}_{-20}\kms$ at a distance of 1.5$^{+1.0}_{-1.0}$ kiloparsec above the disk. The modeling indicates a total mass of $\sim3.9\times10^7\Msun$ for the accreting material at the disk-halo interface on the near side of the M33 disk , with an accretion rate of $\sim2.9\Msunyr$. The high accretion rate and the level of metal-enrichment suggest the inflow is likely to be the fall back of M33 gas from a galactic fountain and/or the gas pulled loosed during a close interaction between M31 and M33. Our study of M33 is the first to unambiguously reveal the existence of a disk-wide, ionized gas inflow beyond the Milky Way, providing a better understanding of gas accretion in the vicinity of a galaxy disk. 

\end{abstract}

%------------------------ Introduction -----------------------------
\section{INTRODUCTION}
\label{sec1}

Chemical studies in the solar neighborhood have shown that the star formation rate (SFR) of the Milky Way (MW; 1.9$\pm$0.4$\Msunyr$; \citealt{Chomiuk11}) has been nearly constant for many billions of years \citep{Chiappini97, Chiappini01}. Since the total molecular gas mass in the disk is only $\sim$ 5$\times$10$^9\Msun$, this reservoir would be exhausted on a significantly shorter timescale without replenishment. Observations of nearby spirals have found the star formation rate (SFR) of a galaxy is tightly related to its molecular gas surface density. This relationship indicates that the molecular gas is consumed at constant efficiency and will be depleted in nearly two Gyrs \citep{Bigiel08}. The common inference drawn from these calculations is that there has to be gas replenishment from an external medium in order to sustain the star-forming activity in a galaxy disk. Analytic studies of galaxies at redshift $z\sim$ 2-3 with high SFRs also show that accretion of additional gas is critical in galaxy formation \citep{Erb08}.  

Observational constraints on the physics of gas accretion onto galaxies are notoriously difficult to obtain. Thus far, the best constraints have been provided by observations of gas in our own MW halo. Absorption- and emission-line studies of the halo reveal complex signatures of multiple-phase gas accretion. \ion{H}{1} high-velocity clouds (HVCs; $|\vlsr|> $ 90$\kms$)\footnote{$\vlsr$ is the velocity in the rest frame of the Local Standard of Rest at the solar circle (LSR).} and intermediate-velocity clouds (IVCs; 30 $<|\vlsr|<$ 90$\kms$) appear to be predominantly falling toward the disk \citep{Wakker01, vanWoerden04, Putman12}. However, estimates of the total accretion rate of \ion{H}{1} complexes find a value below the SFR of the Galaxy even with the inclusion of ionized gas envelopes surrounding \ion{H}{1} complexes \citep{Putman12}. On the other hand, ultraviolet (UV) spectroscopy of bright halo stars and distant quasars (QSOs) reveal ubiquitous strong absorption from ionized gas (\citealt{Sembach03, Collins09, Shull09, Lehner12, Wakker12}). Using the distances to halo stars as upper limits on the foreground absorbing gas cloud distances, \cite{Lehner11b} place a lower limit on the mass accretion rate of 0.4--1.4$\Msunyr$ in the MW halo. This accretion rate is comparable to the SFR of the MW. However, having only the radial velocity component along lines of sight with large physical separations ($\gtrsim2$ kpc) leads to uncertainties regarding the physical arrangement and three-dimensional motions of halo gas. In addition, mass estimates of the MW halo gas could miss $\sim50$\% of the gas which has velocities close to the systemic velocity of the disk, as is found in synthetic observations of a MW-mass galaxy from a cosmological simulation (\citealt{Zheng15, Joung12b}). 

In external galaxies, absorption-line experiments using bright background sources, often QSOs, to study diffuse halo gas have established the existence of an extended, ionized, gaseous halo (aka the circumgalactic medium, or CGM; e.g., \citealt{Lanzetta95, Chen98, Chen01, Prochaska11}). Recent work indicates that the CGM comprises a massive (M$>$10$^{10}\Msun$), spatially extended ($\sim$200 kpc) reservoir of fuel for future star formation and the byproducts of stellar evolution (e.g., \citealt{Tumlinson11, Werk14}). Our neighbour, M31, is also likely to have a massive CGM, as is found by \cite{Lehner15} who studied several QSO sightlines with significant M31-related detections within the virial radii of the galaxy. One major limitation of these QSO absorption-line studies is their inability to detect gas flowing into and out of galaxies. Another unavoidable shortcoming is that such studies are often limited to a single sightline per galaxy (except e.g., \citealt{Keeney13, Chen14, Lehner15}), providing incomplete information on the spatial distribution of gas around galaxies.

Recent cosmological simulations of MW-mass galaxies lend further support to this emerging picture of a dynamic, gaseous baryon cycle within galaxy halos. A hydrodynamical cosmological simulation of a MW-mass galaxy finds that within the virial radius the \ion{H}{1} accretion rate is $\sim0.2\Msunyr$ and the ionized gas accretion rate is $\sim3-5\Msunyr$ (\citealt{Joung12b, Fernandez12}). Broadly speaking, the simulations incorporate a range of feedback prescriptions, yet are unanimous in indicating the presence of ionized gas extending to a galaxy's virial radius (\citealt{Joung12b, Shen12, Oppenheimer12, Ford13, Hummels13, Suresh15, Muratov15}).

To eventually feed star-forming activities in a galaxy's disk, additional gas fuel needs to be obtained from beyond the disk. Cool gas accretion has been observed in star-forming galaxies in \ion{Fe}{2} and \ion{Mg}{2} at redshift $z\sim0.35-1$ \citep{Rubin12}. The gas should be accreted through the disk-halo interface of the galaxy either radially via the edge of the disk \citep{Stewart11} or by directly falling down from above the disk. Therefore, the disk-halo interface serves as a key transition region where the mixing of primordial inflowing gas and metal-enriched outflowing gas occurs. In the MW, it has been observed in multiple phases (e.g., \ion{H}{1}, H$\alpha$, \ion{C}{4}, \ion{Si}{4}, \ion{O}{6}) extending to various heights ranging from a few hundred parsecs to several kiloparsecs (\citealt{Dickey90, Reynolds93, Haffner03, Sembach03, Levine06, Wakker08, Shull09, Putman12}). A hierarchy of increasing temperature and ionization states with increasing scale height has been observed: cold \ion{H}{1} mostly dominates at lower z-heights while warm and hot ionized gas fills most of the volume at greater distances from the plane (\citealt{Dickey90, Gaensler08, Savage09, Putman12}). The densest ionized component of this interface, observed in emission, is called the Reynolds layer or the warm ionized medium layer in the MW (\citealt{Reynolds93, Rand97, Haffner03}).

Kinematically, gas with lagging rotation has been found at the disk-halo interface of both the MW and other spiral galaxies in which the rate of halo gas rotation decreases with z-height (\citealt{Lockman02, Ford10, Saul12, Putman12}). Such a lagging component shows a typical velocity drop-off of $15-30\kms$ kpc$^{-1}$ as observed in deep \ion{H}{1} observations of external galaxies (\citealt{Sancisi01, Fraternali02, Oosterloo07, Heald11}). That is to say, gas at the disk-halo interface has been found to move at velocities very close to the systemic velocity of a galaxy. For the MW observations, the fact that we are residing inside the ISM increases the difficulty in separating the low-velocity disk-halo component from the ISM component. For extragalactic absorption-line observations, low spectral resolutions (typically $\gtrsim50-100\kms$) is a major problem (\citealt{Heckman00, Weiner09, Chen10, Martin12, Rubin12}); it would be challenging, if not impossible, to detect infalling IVC- and HVC-like clouds such as those found in the MW halo if they exist in external galaxies. 

\begin{figure*}[t!]
\includegraphics[width=\textwidth]{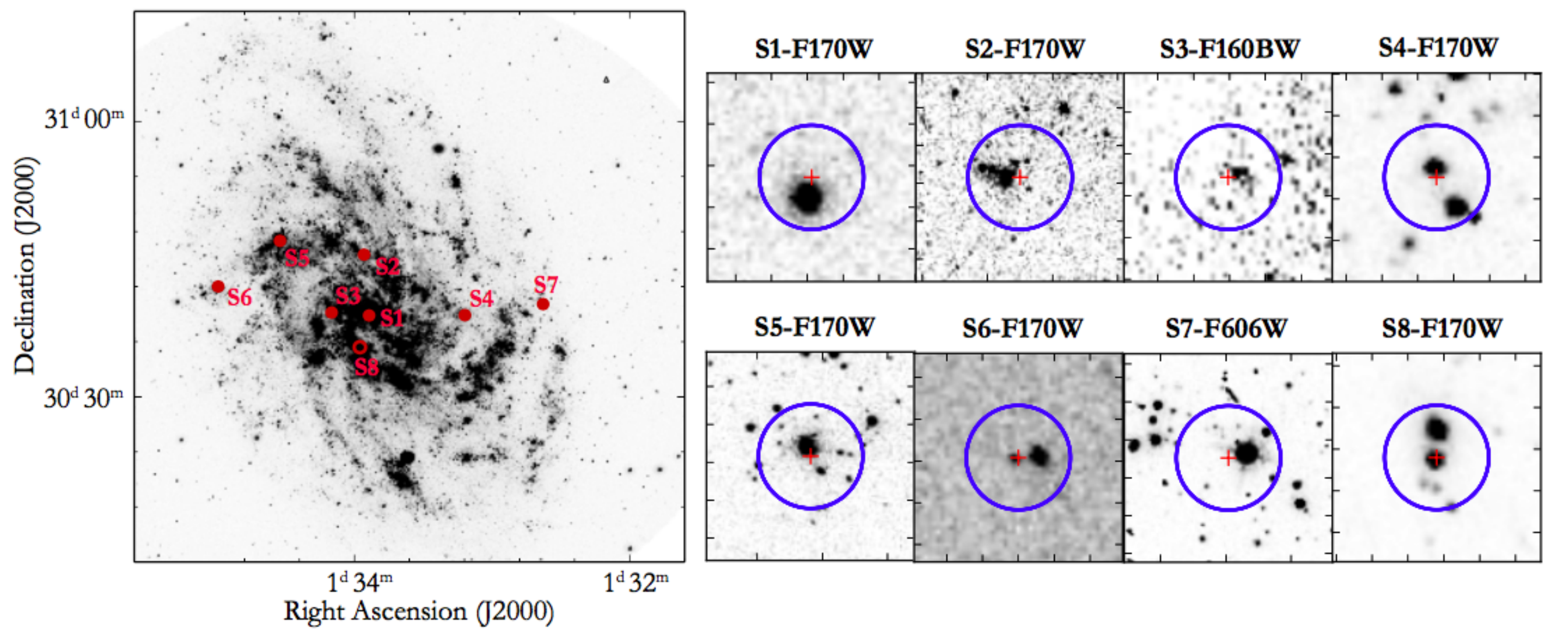}
\caption{Left: The distribution of the target stars (S1-S8) in the disk of M33. The background image is from the {\it Galaxy Evolution Explorer} ({\it GALEX}). S8, indicated by an open circle, is not used in our analysis (see Section \ref{sec2.1}). Right: The number of stars within the COS aperture along each sightline. The background {\it HST}/WFPC images were retrieved from MAST archive. Images taken with F170W filter are used when they exist. The red cross indicates the center of the COS aperture and the blue circle shows its size (2.5" in diameter). At the distance of M33 (840 kpc), 2.5" is $\sim10$ pc. }
\label{fig1}
\end{figure*}

In this paper we investigate the low- and intermediate-velocity gas flows at the disk-halo interface of M33 with medium-resolution spectra (FWHM $\sim14-19\kms$) from the Cosmic Origins Spectrograph (COS) on the {\it Hubble Space Telescope (HST)}. M33 is a late-type Sc galaxy at a distance of 840 kpc \citep{Freedman91} and it moves toward the MW at $\vlsr=-179\kms$ \citep{Corbelli97}. Its proximity makes UV-bright stars in the disk of M33 observable in a reasonable amount of integration time using {\it HST}/COS. The galaxy has an inclination of 56$\degree$ \citep{Paturel03}, ensuring that a fair fraction of the velocity will be projected along the line of sight if vertical gas inflows or outflows exist. The dark matter halo mass of M33 is 5$\times$10$^{10}\Msun$ while its stellar mass is $3-6\times$10$^9\Msun$ \citep{Corbelli03}. It has atomic \ion{H}{1} mass of $1-3\times10^9\Msun$ (\citealt{Deul87, Corbelli97, Putman09, Gratier10}) and molecular mass of $3-4\times10^8\Msun$ (e.g., \citealt{Corbelli03, Gratier10}). M33 is actively forming stars with a SFR of $0.2-0.5\Msunyr$ based on a variety of observations (e.g., \citealt{Engargiola03, Gratier10}). Deep \ion{H}{1} observations and galactic chemical evolution models suggest gas accretion could account for its ongoing star formation (\citealt{Magrini07b, Putman09}). Such gas accretion can be detected with our {\it HST}/COS UV absorption-line studies, as has been shown in observations of the MW ionized HVCs (\citealt{Collins09, Savage09, Shull09, Lehner12}). 

%In the rest frame of the M33 gas disk (see Section \ref{sec2.4}), we define the \it{low-velocity} gas as that with observed velocities (i.e., without an inclination correction) of -40$\lesssim v\lesssim$40$\kms$ and the \it{intermediate-velocity} gas with $v\gtrsim$ +40$\kms$.

Our {\it HST}/COS observations toward M33 provide detailed kinematic information of gas flows on one side of the M33 disk, thus have the great advantage over typical extragalactic observations. In our study, we use UV-bright M33 disk stars as background sources, which helps to unambiguously determine the relative gas motions with respect to the disk. In the rest frame of the M33 gas disk (see Section \ref{sec2.4}), we define the {\it low-velocity} gas as that with observed velocities (i.e., without an inclination correction) of $-40\lesssim v_{\rm d}\lesssim+40\kms$ and the {\it intermediate-velocity} gas with $v_{\rm d}\gtrsim+40\kms$; $v_{\rm d}$ shall be defined in Section \ref{sec2.4}. By observing multiple sightlines toward different locations of the disk, we are able to differentiate between various accretion models, obtain a better understanding of how gas is flowing, and assess the overall accretion rate. Similar techniques to study disk-wide gas kinematics have also been applied to the observations of multi-phase gas associated with the Large Magellanic Cloud (LMC; \citealt{Howk02, Danforth06, Lehner09, Pathak11, Barger16}) and the Small Magellanic Cloud (SMC; \citealt{Hoopes02}). 

The article is structured as follows. Section \ref{sec2} outlines the sample selection criteria and spectral reduction processes (including continuum fitting and Voigt-profile fitting), and addresses the CalCOS wavelength calibration uncertainty. We calculate the systemic velocity of M33's ISM along each sightline, define the $v_{\rm d}$ frame (the rest frame of M33's ISM), and estimate M33's hydrogen content along each sightline in this section. Section \ref{sec3} describes the ion properties. We show that the detected absorbers are indeed associated with M33 in Section \ref{sec4}. In Section \ref{sec5} we perform kinematic modeling of the consistently redshifted \ion{Si}{4} absorption lines, and in Section \ref{sec6} address the metal enrichment level of detected gas inflow using our absorption-line measurements and photoionization modeling. In Section \ref{sec7} we discuss the accretion rate, the origin(s) of the inflowing gas, and the possibility of galactic outflows from M33. We conclude with a summary of our main findings in Section \ref{sec8}.

%\input t1.tex
%%%%%%%%%%%%%%%%%%%%%%%%%%%%%%
\begin{table*}
\renewcommand{\arraystretch}{0.9}
\caption{M33 Sample}
\footnotesize
\begin{center}
\begin{tabular}{cccccccccc}
\hline
\hline
Name & S-ID\tablenotemark{a} & RA (J2000) & DEC (J2000) & ${v_{\rm rot}}$\tablenotemark{b} & $R$\tablenotemark{c} & $R_{\rm G}$\tablenotemark{c} & Exp. T. & Sp. Type\tablenotemark{f} & Note \\
 & & (hh mm ss) & (dd mm ss) & (${\rm km\ s^{-1}}$) & (kpc) & (kpc) & (s) & & \\
\hline
M33-UIT-236 & S1 & 01 33 53.60 & +30 38 51.60 & -177.0 & 0.2 & 0.4  & 10,651 & Ofpe/WN9 & WR\\ 
M33-FUV-350 & S2 & 01 33 56.00 & +30 45 31.00 & -233.7 & 1.5 & 1.5  & 11,172 & A8-F0Ia  & Late A type supergiant\tablenotemark{d}\\
M33-FUV-444 & S3 & 01 34 09.90 & +30 39 11.00 & -194.7 & 1.2 & 1.7  & 11,200 & O6III    & Star\\
NGC592	    & S4 & 01 33 12.30 & +30 38 49.00 & -160.1 & 2.4 & 3.4  & 11,168 & WN3+neb  & \ion{H}{2} region; WR2\tablenotemark{e}\\
NGC604      & S5 & 01 34 32.50 & +30 47 04.00 & -240.3 & 3.1 & 3.6  & 11,221 & WN10     & \ion{H}{2} region; WR12\tablenotemark{e}\\
M33-OB-88-7 & S6 & 01 34 59.40 & +30 42 01.10 & -210.6 & 4.2 & 5.9  & 11,156 & O8Iaf    & blue supergiant\\
M33-FUV-016 & S7 & 01 32 37.70 & +30 40 06.00 & -157.6 & 4.5 & 6.6  & 11,220 & Ofpe/WN9 & WR\\    
\hline 
M33-OB-2-4  & S8 & 01 33 58.70 & +30 35 27.00 & -147.0 & 1.1 & 1.5  & 11,222 & Ofpe/WN9 & WR + B supergiant \\
\hline
\hline
\end{tabular}
\end{center}
\tablenotetext{a}{S-ID is in the order of increasing $R_{\rm G}$ with S1 being the closest and S7 the most distant. S8 does not follow this convention since it is not used in our analysis. See Section \ref{sec2.1} for the explanation.}
\tablenotetext{b}{${v_{\rm rot}}$: the rotation velocity of the gas disk of M33 at the position of a given sightline. See Section \ref{sec2.4} for the derivation of $v_{\rm rot}$. }
\tablenotetext{c}{$R$: the projected galactocentric distance. $R_{\rm G}$: the inclination-corrected galactocentric distance. }
\tablenotetext{d}{\cite{Humphreys13}. }
\tablenotetext{e}{Several WR stars are reported in these two \ion{H}{2} regions; based on coordinate coincidence, we find sightline S4 is likely pointing at NGC592-WR2 and S5 at NGC604-WR12 as identified in \cite{Drissen08}.}
\tablenotetext{f}{Spectral type. S1, S4, S5, S7: \cite{Neugent11}; S2: \cite{Humphreys13}; S3, S6, S8: \cite{Massey06}.}
\label{tb1}
\end{table*}

%%%%%%%%%%%%%%%%%%%%%%%%

%------------------------ Method -----------------------------
\section{OBSERVATIONS AND MEASUREMENTS}
\label{sec2}

\subsection{Sample Selection and COS Spectroscopy}
\label{sec2.1}

We targeted UV-bright stars in massive OB associations in the disk of M33. Each star was required to have {\it Far Ultraviolet Spectroscopic Explorer} ({\it FUSE}) spectra available in the Mikulski Archive for Space Telescopes (MAST), and its UV continuum flux at 1300 \AA\ was required to be $>5\times10^{-15}$ erg s$^{-1}$ cm$^{-2}$ \AA$^{-1}$ to achieve a S/N of $15-20$. These criteria limited us to only 14 stars. Among all the candidates, we selected seven stars evenly distributed across the disk of M33. This helps sample the gas disk rotation at different locations and galactocentric radii, and thus distinguish among different accretion models (see Section \ref{sec5}).

The seven targeted stars were observed in $2014-2016$ during {\it HST} Cycle 22 (proposal ID: 13706) using the G130M FUV grating with a central wavelength of 1291 \AA\ and coverage from 1134 \AA\ to 1431 \AA. The resolving power of COS/G130M centering at 1291 \AA\ is $R=16,000-20,000$, resulting in a spectral resolution (FWHM) of $\sim14-19\kms$ (COS data handbook; \citealt{Fox15}). The observations were conducted using the FUV XDL detector under TIME-TAG mode with the Primary Science Aperture (2.5"; $\sim$10 pc at the distance of M33). Each star was observed with four exposures with a total integration time of $\sim11,000$ s (see Table \ref{tb1}). We retrieved the calibrated and co-added spectra from MAST, which have been processed by the standard CalCOS (version 3.0) pipeline. To supplement our sample, we also retrieved the {\it HST}/COS spectrum of M33-UIT-236 -- a Wolf-Rayet (WR) star near the disk center. The observation of M33-UIT-236 was carried out in 2011 under the {\it HST} Cycle 18 COS Guaranteed Time observation program (proposal ID: 12026), with a similar spectrograph setting as ours \citep{Welsh13}. 

\begin{figure*}[t]
    \centering
    \includegraphics[width=\textwidth]{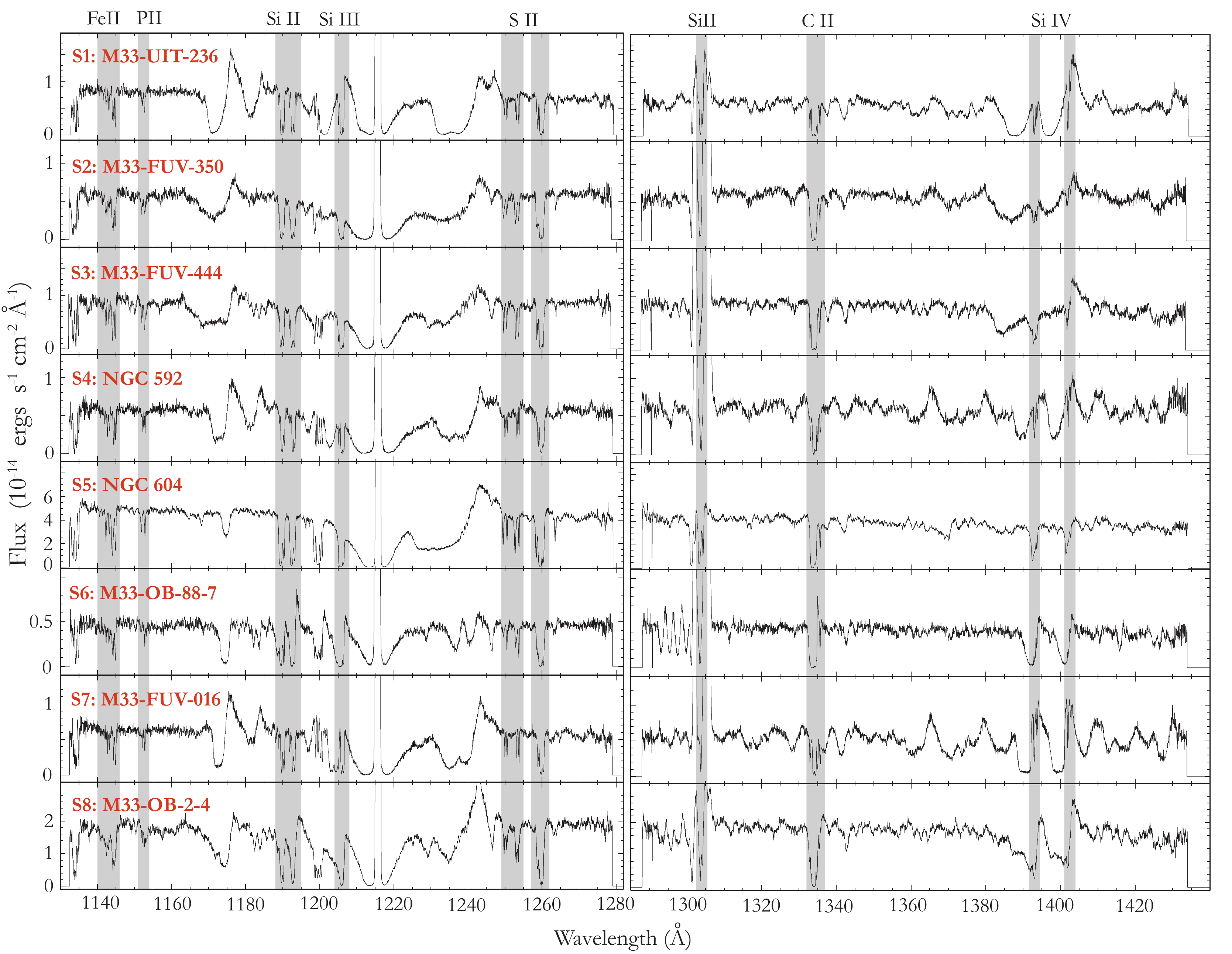}
    \caption{CalCOS co-added stellar spectra. We bin the spectra by five pixels for better illustration. Interstellar lines are highlighted in gray, including \ion{Fe}{2} $\lambda\lambda\lambda$1142, 1143, 1144, \ion{P}{2} $\lambda$1152, \ion{Si}{2} $\lambda\lambda\lambda$1190, 1193, 1304, \ion{Si}{3} $\lambda$1206, \ion{S}{2} $\lambda\lambda$1250, 1253, \ion{C}{2} $\lambda$1334, \ion{Si}{4} $\lambda\lambda$1393, 1402. Note that \ion{Si}{2} $\lambda$1304 is dominated by the airglow emission in this figure; for the continuum-normalized spectra shown in the following figures, we use the night-only data retrieved from the original observation for this line. We describe the night-only data reduction in detail in Zheng et al. 2017, (in prep). Due to the wide wavelength span of the spectra, M33 and MW components of the same line are usually very close to each other in this figure. For a better illustration of line sub-structures, please see Fig \ref{figc3}-\ref{figc6} . We do not use sightline S8 in our analysis since its interstellar lines are strongly affected by stellar activities and, the spectrum is likely a combination of a WR star and a blue supergiant (see Section \ref{sec2.1}).}
\label{fig2}
\end{figure*}

Table \ref{tb1} lists the position (RA, DEC), rotation velocity of the gas disk at the position of a given sightline $v_{\rm rot}$ (Section \ref{sec2.4}), projected galactocentric distance $R$ and inclination-corrected galactocentric distance $R_{\rm G}$, total exposure time, and spectral type. For clarity, we assign to each sightline an ID S1--S7 in the order of increasing $R_{\rm G}$, with S1 being the closest to the disk center and S7 the most distant. S8 (M33-OB-2-4) does not follow this rule since it is not included in our analysis (see below for explanation). 

Emission- and absorption-line studies in the literature show that our targets have active stellar winds and are fast rotators (e.g., \citealt{Humphreys13, Drissen08, Tenorio00}). Such activity generates stellar and photospheric lines that can be easily distinguished from interstellar lines because of their unique line shapes (e.g., P Cygni profiles) and large line widths. We describe the treatment of stellar lines and the normalization of stellar continua in Section \ref{sec2.3} and Appendix \ref{appC}. 

The left panel of Fig \ref{fig1} shows the locations of S1-S8 in the disk of M33. They are mostly associated with spiral arms and/or \ion{H}{2} regions. For each sightline, we retrieve the {\it HST}/F170W\footnote{Other {\it HST} filters are chosen when F170W is not available.} images from MAST archive and, examine the number of stars within the {\it HST}/COS aperture. The right panels of Fig \ref{fig1} show that most sightlines have only one UV-bright star dominating the COS aperture except for S4 and S8. 

S4 is associated with a giant \ion{H}{2} region that has several WR stars identified in the field. Based on coordinate coincidence, one of the two stars in S4 is likely the NGC592-WR2 identified in \cite{Drissen08}. Since the {\it HST}/COS spectrum of S4 is similar to others shown in Fig \ref{fig2} and \ref{fig3}, the two stars within the aperture are not likely to cause significant blending and smearing of the spectral lines. We therefore decided to keep this sightline in our analysis.

Along sightline S8, two bright stars are present within the COS aperture. Spectroscopic analysis indicates that one is a WN-type WR star and the other is a B supergiant ( \citealt{Massey95, Massey96}). Our {\it HST}/COS spectra suggests a combination of the two stars, as shown in the bottom panel in Fig \ref{fig2}. The interstellar lines of interest along sightline S8, as highlighted in gray, are clearly dominated by stellar features such as P Cygni profiles. This makes the continuum normalization and Voigt-profile fitting highly unreliable. Therefore, we exclude this sightline from our analysis. Note that this does not represent a non-detection in our sample. Our final target list includes seven sightlines that are listed as S1--S7 in Table \ref{tb1}.  

In addition, we retrieved the {\it FUSE} spectra for our sightlines from MAST. The spectra span a wavelength range of $905-1187$ \AA\ within which an important absorption line \ion{O}{6} $\lambda$1032 lies. Since \ion{O}{6} is not the focus of this work and we only use it for a comparison with \ion{Si}{4}, a simpler spectral reduction was performed. The stellar continuum within $\pm$1000$\kms$ of 1032 \AA\ was normalized using first- and second-order polynomial functions. We did not attempt to run Voigt-profile decomposition for \ion{O}{6} $\lambda$1032 given the low S/N of the {\it FUSE} data. The normalized \ion{O}{6} absorption lines and relevant discussion are presented in Section \ref{sec6.1}. 

All the {\it HST},  {\it FUSE} and {\it GALEX} data used in this paper can be found in MAST: \dataset[10.17909/T9FG6R]{http://dx.doi.org/10.17909/T9FG6R}.

\subsection{Wavelength Calibration and Spectral Co-addition}
\label{sec2.2}

As mentioned in Section \ref{sec2.1}, each sightline was observed with four exposures which produce four spectra that need to be co-added. The standard CalCOS pipeline provides data reduction for spectral co-addition and wavelength calibration with an uncertainty of nearly one resolution element. We show the original CalCOS-processed stellar spectra in Fig \ref{fig2}. In the following work, we process and present the spectra in their original resolution; we do not perform any binning to the spectra unless otherwise specified. 

Several authors have pointed out that problems may arise with CalCOS wavelength calibration and spectral co-addition, and thus have written their own pipelines to process {\it HST}/COS spectra in order to minimize the uncertainties. To justify that CalCOS products are reliable for our scientific analysis, we used three other pipelines to calibrate and co-add the original spectra and compared the results with those from CalCOS. The three pipelines we investigated are: (1) an IDL routine x1d\_coadd.pro by \cite{Danforth10}, (2) a spectral co-add code by \cite{Wakker15}, and (3) the PyCOS pipeline by \cite{Liang14} (\& private communication). We discuss the details of these pipelines and compare them with CalCOS in Appendix \ref{appB}. All the coadded spectra processed by these three methods can be found in \citet[][Dataset: \url{https://doi.org/10.5281/zenodo.168580}]{Zheng16}

In brief, our investigation shows consistency between CalCOS products and those from other pipelines. A couple of lines using the method of \cite{Wakker15} have minor wavelength shifts with respect to the CalCOS spectra but all within one resolution element. We note that such good agreement between CalCOS and the other three pipelines is mainly due to the straightforward setup of our observations. For each of our sightlines, the observation was completed with four exposures in one single visit and the spectra were taken under the same setting. The background stars are all UV-bright to ensure high S/N. Thus the possibility of spectral miss-alignment is largely reduced. CalCOS pipeline is most likely to become problematic in situations where faint QSO observations and multiple spectra are obtained at different epochs. We proceed with our analysis using the CalCOS co-added spectra that we have tested to be scientifically reliable by direct comparisons with three other different methods.

\subsection{Line Identification, Continuum Fitting and Voigt-Profile Fitting}
\label{sec2.3}

Our observations include interstellar absorption lines of \ion{S}{2} $\lambda\lambda$1250, 1253\footnote{\ion{S}{2} lines are in fact triplet: 1250 \AA, 1253 \AA, and 1259 \AA. However, \ion{S}{2} $\lambda$1259 is blended with a $\vlsr\sim-350\kms$ component of \ion{Si}{2} $\lambda$1260 that we will discuss in Section \ref{sec7.3} and Appendix \ref{appA}. We do not include \ion{S}{2} $\lambda$1259 in our analysis. We also exclude \ion{Si}{2} $\lambda$1260 and only use \ion{Si}{2} $\lambda\lambda\lambda$1190, 1193, 1304.}, \ion{Fe}{2} $\lambda\lambda\lambda$1142, 1143, 1144, \ion{P}{2} $\lambda$1152, \ion{Si}{4} $\lambda\lambda$1393, 1402, \ion{Si}{2} $\lambda\lambda\lambda$1190, 1193, 1304\footnote{In each of our COS spectra, \ion{Si}{2} $\lambda$1304 is strongly affected by nearby airglow emission \ion{O}{1} $\lambda\lambda$1302, 1304 due to oxygen atoms in the exosphere of the Earth. For this line, we retrieve night-only photons from the spectra, the process of which is described in detail in Zheng et al., 2017, (in prep.). }, \ion{Si}{3} $\lambda$1206, and \ion{C}{2} $\lambda$1334. In Fig \ref{fig2}, we highlight these lines in gray. Due to the broad wavelength span, M33 and MW absorption components of the same line are very close to each other in this figure. Please see Fig \ref{figc3}--\ref{figc6} for a better illustration of the line structures.

Line identification is made according to wavelength coincidence at the rest frame of the gas disk at the position of the corresponding sightline. For each absorption line, we confirm its detection in each of the four exposures to ensure that it is not an artifact due to spectral co-addition or fixed pattern noise (COS data handbook; \citealt{Fox15}). Our spectra also cover \ion{N}{5} $\lambda\lambda$1238, 1242, but we do not detect related absorption features with reliable significance in any of our sightlines.  

The narrow interstellar absorption lines of interest are found superimposed on stellar features that have much broader line widths and commonly have P Cygni profiles. We treat the stellar features as continua such that interstellar line profiles can be normalized. First, we model the continua using Legendre polynomials within $\pm$1000$\kms$ from the centroids of the lines of interest. The fitting is evaluated such that the polynomial order is kept as low as possible while the reduced $\chi^2$ ($\chi^2_{\rm red}$) is minimized to $\sim1.0$. For ions (\ion{Fe}{2}, \ion{S}{2}, \ion{Si}{4}) that have multiple transition lines available and the lines are not saturated, we use their oscillator strengths $f$ to scale their normalized profiles and compare the $f$-scaled line shapes. This $f$-scale matching helps to simultaneously constrain the continuum fitting of different transition lines of the same ion, and ensures that we have filtered out the stellar features plaguing the interstellar lines of interest at different wavelengths. 

We note that unresolved saturation may exist in the non-saturated lines even though their minimum fluxes have not yet reached zero, which will make the $f$-scale matching not valid. To evaluate the level of unresolved saturation, we perform another independent continuum fitting for each non-saturated line of \ion{S}{2} and \ion{Si}{4}. We use a different pipeline -- the linetools package\citep{linetools}\footnote{\href{url}{https://github.com/linetools/linetools}}, which applies Akima Spline to regenerate the continua of non-absorption region of each line. We compare the column densities of \ion{S}{2} (\ion{Si}{4}) derived from $\lambda$1250 and $\lambda$1253 ($\lambda$1393 and $\lambda$1402), and find that the values agree well. For \ion{S}{2} lines, the log $N_{\rm SII}$ derived from $\lambda1253$ are generally consistent with those from $\lambda1250$ within $\sim0.05$ dex; only in S2 and S7 we find an offset of $\sim0.1-0.2$ dex. For \ion{Si}{4} lines, the $N_{\rm SiIV}$ derived from $\lambda$1393 and $\lambda$1402 are also consistent within 0.05 dex. Therefore, we conclude that the unresolved saturation is well controlled in our COS data and only along a couple of sightlines (e.g., S7-\ion{S}{2}) the unresolved saturation may affect the column density estimates by up to 0.2 dex. We refer the reader to Fig \ref{figc3}--\ref{figc6} for a close inspection on the continuum fitting and $f$-scaled profile matching of each line. 

%\input t2.tex
%%%%%%%%%%%%%%%%%%%%%%%%
\begingroup
\renewcommand\arraystretch{0.96}
\begin{ThreePartTable}
\begin{longtable}{ccccc}
\caption{Voigt Profile Fitting\label{tb2}}  \\
\hline
\hline
S-ID & $v_{\rm d}$ $^a$ & $b$ $^a$ & log $N$ $^a$& $\chi^2_{\rm red}$ $^a$ \\
 & ($\kms$) & ($\kms$) & (${\rm cm^{-2}}$) & \\
\hline
\endfirsthead
\multicolumn{5}{c}
%{\tablename\ \thetable\ -- \textit{Continued}} \\
{\textbf{Table 2.} -- \textit{Continued}} \\
\hline
S-ID & $v_{\rm d}$ $^a$ & $b$ $^a$ & log $N$ $^a$& $\chi^2_{\rm red}$ $^a$ \\
\hline
\endhead
\endfoot
\multicolumn{5}{>{\footnotesize}c}{\ion{S}{2}: 1250.58 \AA, 1253.81 \AA} \\
\hline
S1  & -15.0$\pm$ 5.7 &  41.3$\pm$ 4.5 & 15.07$\pm$0.07  & 0.94 \\ 
    &  20.0$\pm$ 1.0 &  17.0$\pm$ 2.3 & 14.87$\pm$0.10  &      \\ 
S2  & -10.7$\pm$ 5.5 &  35.0$\pm$ 4.9 & 15.18$\pm$0.16  & 0.62 \\ 
	&  51.2$\pm$17.6 &  52.6$\pm$14.7 & 15.15$\pm$0.18  &      \\ 
S3	& -18.6$\pm$ 0.5 &  30.2$\pm$ 0.7 & 15.34$\pm$0.01  & 0.92 \\ 
S4	&  29.2$\pm$ 0.8 &  31.5$\pm$ 1.3 & 15.24$\pm$0.02  & 0.71 \\ 
S5	& -20.6$\pm$ 9.5 &  52.1$\pm$ 3.8 & 15.14$\pm$0.18  & 0.92 \\ 
    &  -4.0$\pm$ 1.9 &  33.5$\pm$ 3.4 & 15.13$\pm$0.17  & 	   \\ 
    &  52.8$\pm$ 1.4 &  13.3$\pm$ 3.4 & 14.01$\pm$0.18  & 	   \\ 
S6	&  -2.3$\pm$ 4.5 &  39.4$\pm$ 7.6 & 15.10$\pm$0.08  & 0.65 \\ 
    &  68.7$\pm$10.6 &  29.3$\pm$13.6 & 14.54$\pm$0.25  & 	   \\ 
S7	&  15.2$\pm$ 0.7 &  18.3$\pm$ 1.0 & 15.02$\pm$0.02  & 0.95 \\ 
\hline
\multicolumn{5}{>{\footnotesize}c}{\ion{P}{2}: 1152.82 \AA} \\
\hline
S1	& -37.0$\pm$29.8 &  26.4$\pm$23.7 & 13.10$\pm$0.86 & 0.64 \\ 
    &   9.1$\pm$24.0 &  34.1$\pm$20.2 & 13.46$\pm$0.38 &      \\
S2	& -10.9$\pm$ 7.6 &  37.5$\pm$ 9.4 & 13.66$\pm$0.11 & 0.54 \\
	&  69.9$\pm$15.1 &  38.3$\pm$28.5 & 13.31$\pm$0.28 &      \\ 
S3	& -18.8$\pm$ 2.9 &  34.0$\pm$ 4.0 & 13.62$\pm$0.04 & 0.91 \\
S4	&  32.3$\pm$ 3.4 &  37.9$\pm$ 4.7 & 13.77$\pm$0.05 & 0.57 \\
S5	&   2.6$\pm$ 1.4 &  40.9$\pm$ 2.1 & 13.69$\pm$0.02 & 1.00 \\
S6	& [-80, 80]$^b$ & - & $\leq$13.3$^b$ & \\
S7	&  19.9$\pm$ 3.3 &  25.9$\pm$ 4.7 & 13.47$\pm$0.06 & 0.72 \\
\hline
\multicolumn{5}{>{\footnotesize}c}{\ion{Fe}{2}: 1142.37 \AA, 1143.23 \AA, 1144.94 \AA} \\
\hline
S1 	& -41.1$\pm$9.8 &  31.2$\pm$7.4 & 14.34$\pm$0.20 & 0.81 \\
    &  10.7$\pm$10.4&  34.6$\pm$8.1 & 14.42$\pm$0.16        \\
S2	&   1.0$\pm$6.2 &  39.2$\pm$6.4 & 14.56$\pm$0.08 & 0.76 \\
	&  68.3$\pm$4.8 &  31.0$\pm$4.9 & 14.51$\pm$0.08 &      \\
S3	& -25.6$\pm$3.8 &  18.5$\pm$9.7 & 13.96$\pm$0.39 & 0.86 \\
    & -23.4$\pm$2.6 &  46.2$\pm$6.7 & 14.52$\pm$0.11 &      \\
S4	&  42.2$\pm$2.9 &  24.2$\pm$4.0 & 14.31$\pm$0.06 & 0.70 \\
S5	& -39.9$\pm$2.5 &  12.9$\pm$5.9 & 13.37$\pm$0.29 & 1.11 \\
    &   8.9$\pm$1.8 &  40.4$\pm$4.2 & 14.63$\pm$0.04 &      \\
    &  74.6$\pm$4.8 &  32.5$\pm$3.9 & 14.11$\pm$0.09 &      \\
S6	&  -7.2$\pm$11.8 & 30.0$\pm$10.8& 14.39$\pm$0.21 & 0.80  	\\
	&  39.5$\pm$10.5 & 23.7$\pm$10.4& 14.23$\pm$0.30 &  \\
	&  96.0$\pm$ 8.1 & 14.2$\pm$12.6& 13.42$\pm$0.27 &   	\\
S7	&   9.9$\pm$1.3 &  18.6$\pm$1.8 & 14.34$\pm$0.04 & 0.87 \\
\hline
\multicolumn{5}{>{\footnotesize}c}{\ion{Si}{4}: 1393.76 \AA, 1402.77 \AA} \\
\hline
S1	&  3.0$\pm$0.8 &  32.8$\pm$ 1.0 & 14.17$\pm$0.01 & 0.84 \\ 
	& *67.9$\pm$1.8$^c$ &  22.3$\pm$ 2.0 & 13.38$\pm$0.04 & 	 \\ 
S2	& 10.4$\pm$20.3 &  44.8$\pm$14.6 & 13.33$\pm$0.54 & 0.69 \\ 
	& *75.6$\pm$32.4 &  60.9$\pm$21.7 & 13.57$\pm$0.31 & 	 \\ 
S3	&-14.9$\pm$2.3 &  44.3$\pm$ 3.2 & 13.67$\pm$0.03 & 0.96 \\ 
	& *75.9$\pm$3.1 &  34.2$\pm$ 4.4 & 13.35$\pm$0.05 &	 \\ 
S4	&-29.3$\pm$3.5 &  14.7$\pm$ 3.8 & 12.94$\pm$0.13 & 0.65 \\ 
	&  8.4$\pm$2.3 &  20.4$\pm$ 4.5 & 13.43$\pm$0.14 & 	 \\ 
	& *50.8$\pm$13.3 &  34.5$\pm$11.8 & 13.19$\pm$0.20 & 	 \\ 
S5	&-35.6$\pm$2.0 &  34.1$\pm$ 2.2 & 13.47$\pm$0.03 & 0.92 \\ 
	&  5.0$\pm$1.5 &  12.9$\pm$ 2.8 & 12.78$\pm$0.12 &      \\
	& *39.1$\pm$2.8 &  11.3$\pm$ 4.5 & 12.29$\pm$0.12 & 	 \\ 
S6	& -8.9$\pm$5.4  &  19.2$\pm$ 7.4 & 13.20$\pm$0.14 & 0.35 \\ 
	& *70.6$\pm$4.0  &  17.5$\pm$ 5.7 & 13.25$\pm$0.12 &      \\
	& *106.2$\pm$3.9  &  14.5$\pm$ 4.4 & 13.13$\pm$0.14 &      \\
S7	&-19.8$\pm$2.0 &  19.8$\pm$ 1.9 & 13.52$\pm$0.05 & 0.70 \\ 
	& 13.8$\pm$2.0 &  15.6$\pm$ 3.4 & 13.25$\pm$0.14 & 	 \\ 
	& *48.2$\pm$15.0 &  30.6$\pm$15.4 & 12.80$\pm$0.28 & 	 \\ 
\hline
\hline
\end{longtable}
\begin{samepage}
  \begin{tablenotes}
  \footnotesize
  \item[]{Note: }
  \item[a]{(v$_{\rm d}$, $b$, log $N$) are the centroid velocity, Doppler width, and logarithmic column density of the Voigt-profile fits. $\chi^2_{\rm red}$ is the reduced $\chi^2$ of the fit that is calculated within $[-150, 150]\kms$ of the respective spectral line. }
  \item[b]{No detection of \ion{P}{2} along this sightline. We derive a 3$\sigma$ upper limit by integrating the spectra within $[-80, 80]\kms$.}
  \item[c]{The * sign indicates the intermediate-velocity/non-disk \ion{S}{4} component. See Section \ref{sec3} for description.}
  \end{tablenotes}
\end{samepage}
\end{ThreePartTable}
\endgroup
%%%%%%%%%%%Table 2 end%%%%%%%%%%%%%%%%%%

After continuum normalization, we perform Voigt-profile fitting to decompose the interstellar absorption lines by taking into account both the profile shapes and line saturation. By doing so, we resolve the underlying kinematic components of detected absorption lines. We use the same method as described in \cite{Tumlinson13}. First, visual inspection is required to determine the number of kinematic components and unrelated nuisance absorption lines (e.g., absorption due to MW's ISM). Using the MPFIT\footnote{\href{url}{http://purl.com/net/mpfit}}  software \citep{Markwardt09}, we optimize the fit of each component and derive the best-fit column density log $N$, Doppler width $b$ and centroid velocity $v_{\rm d}$ for each component. $v_{\rm d}$ is defined as the velocity in the rest frame of M33's ISM at the position of each sightline; we will define this $v_{\rm d}$ in Section \ref{sec2.4}. The errors for the fit are computed from parameter covariance matrices by MPFIT. Specifically, if an ion has multiple transitions, such as \ion{Si}{4} $\lambda\lambda$1393, 1402, the same number of kinematic components and the same set of parameters (log $N$, $b$, $v_{\rm d}$) will be simultaneously applied to all observed lines. The parameters are eventually determined when the values converge after iteratively fitting. This approach reduces the risk that visual inspection could be biased by artifacts at certain wavelengths. For this simultaneous multi-line fitting, our procedure also takes into account potential velocity shifts and geometric distortion of line profiles at different locations on the grating. The modeled Voigt profiles are convolved with the COS line-spread-function that is given at the nearest observed wavelength grid point in the compilation \citep{Ghavamian09}. We refer the reader to \cite{Tumlinson13} for a detailed discussion on the methodology and the profile fitting algorithm. We tabulate the Voigt-profile fitting results in Table \ref{tb2} and discuss the interpretation in Section \ref{sec3}.

%While Voigt-profile fitting finds ($v_{\rm d}$, $b$, log $N$) for individual kinematic components of each absorption line, it is subject to underlying unresolved structures and potential line saturation that is not recognized. To test the robustness of the Voigt-profile fitting, we additionally use the apparent optical depth (AOD) method (\citealt{Savage91}; \citealt{Savage96}) to derive the total column densities of the same absorption lines. The AOD method converts the absorption line profiles into apparent column densities pixel-by-pixel without prior knowledge of the lines' velocity structures (see Appendex \ref{appA} for a detailed description of this method). Thus, if the line of interest is not saturated, the total column density from AOD method and that from the Voigt-profile fitting should match. We use AOD method to calculate log $N$ of each ion (\ion{Fe}{2}, \ion{P}{2}, \ion{S}{2}, and \ion{Si}{4}) by integrating the corresponding spectrum from $v_{\rm d}\sim-100\kms$ to $\sim+100\kms$. Then we compare the value with that from Voigt-profile fitting that is summed over all the fitted components for each sightline as listed in Table \ref{tb2}. We find that the log $N$ difference between these two methods is within 1 $\sigma$ column density error quoted in Table \ref{tb2}. Thus, we conclude that the saturation of \ion{Fe}{2}, \ion{S}{2}, \ion{P}{2}, and \ion{Si}{4} lines is negligible and, our Voigt-profile fitting correctly reflect the velocity structures of these non-saturated lines. 

The above multi-component and/or multi-line fitting is performed only for ions with non-saturated absorption lines, including those of \ion{P}{2}, \ion{S}{2}, \ion{Fe}{2}, and \ion{Si}{4}. For ions with saturated and blended lines -- i.e., \ion{C}{2}, \ion{Si}{2}, and \ion{Si}{3} -- we do not attempt to recover the underlying kinematic components. As shown in Fig \ref{figc7}, \ion{Si}{2} $\lambda$1304 has the lightest saturation among these lines, even so, the core of the line (M33's ISM) can be still found saturated. Therefore, we only derive the column densities of these ions by integrating the spectra over the M33 velocity ranges using the AOD method (\citealt{Savage91, Savage96}). Since the lines are saturated, these derived values should be lower limits. We describe the AOD method and show the results in Appendix \ref{appA}. A full display of the detected absorption lines along each sightline can be found in Fig \ref{figc7}. In Appendix \ref{appC}, we discuss in detail the continuum- and Voigt-profile fitting of specific lines. 

% In the following sections, we focus on the properties and interpretation of ions \ion{Si}{4}, \ion{S}{2}, \ion{P}{2}, and \ion{Fe}{2} that have non-saturated absorption lines. 
%\input t2.tex

\subsection{Rotation Velocity of M33's ISM $v_{\rm rot}$ at the Position of Each Sightline}
\label{sec2.4}

Throughout this paper, we use a velocity frame, $v_{\rm d}$, which is defined as the velocity in the rest frame of M33's ISM at the position of each sightline: $v_{\rm d}\equiv\vlsr$-$v_{\rm rot}$, where $\vlsr$ is a spectral line's LSR velocity with respect to the Sun and $v_{\rm rot}$ is the rotation velocity of the M33 gas disk at the position of each sightline in the LSR frame. We emphasize that the gas kinematics we investigate is with respect to the M33'ISM, the rotation of which will be determined from \ion{H}{1} 21-cm spectra as shown in the following paragraphs. The exact positions of the background stars in the gas disk do not change our determination of gas inflows and outflows.  

\begin{figure*}[t]
\includegraphics[width=\textwidth]{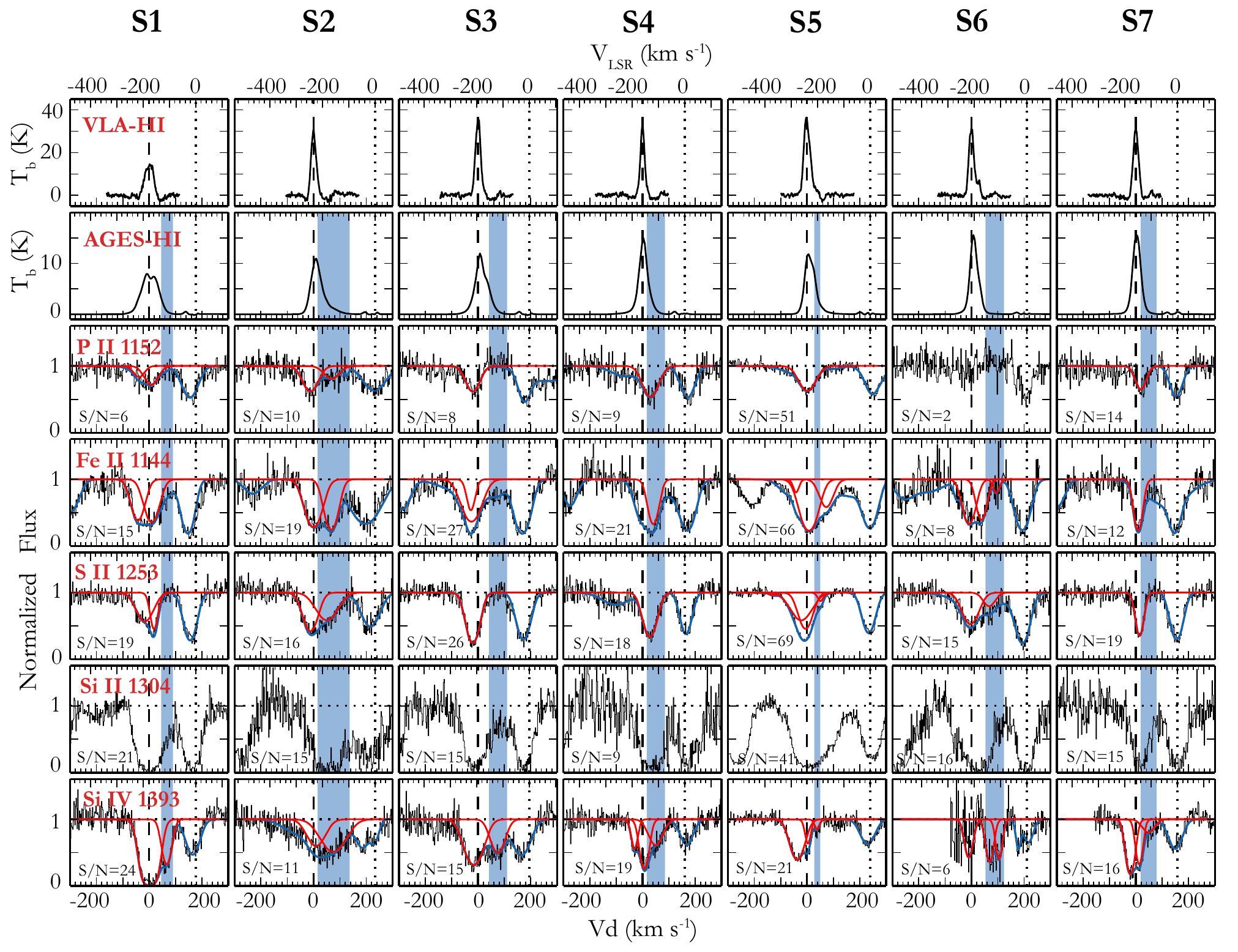}
\caption{The first and second row show the \ion{H}{1} 21-cm spectra from the VLA and the AGES, respectively. Note that the brightness temperature $T_{\rm b}$ is on a different scale. The bottom four rows show the continuum-normalized {\it HST}/COS spectra of \ion{P}{2} $\lambda$1152, \ion{Fe}{2} $\lambda$1144, \ion{S}{2} $\lambda$1253, and \ion{Si}{4} $\lambda$1393. Velocity on the bottom x-axis, $v_{\rm d}$, is with respect to the M33 gas disk rotation velocity at the position of each sightline (see Section \ref{sec2.4}); velocity on the top x-axis in the LSR frame at solar circle. The absorption lines from M33's ISM with $v_{\rm d}=0\kms$ are shown by dashed lines, while those from MW's ISM with $\vlsr=0\kms$ are indicated by dotted lines. The red curves in each panel indicate the Voigt-profile fitted components, while the underlying blue curves show the overall fitted profiles including additional nuisance components. For example, the M33 \ion{Fe}{2} $\lambda$1144 along S4 is blended with the MW \ion{Fe}{2} $\lambda$1143, thus a nuisance is added at $v_{\rm d}\sim0\kms$. The blue shaded areas highlight the redshifted non-disk \ion{Si}{4} components (Section \ref{sec3}). The S/N ratio of each line is shown in the bottom left corner, which is measured locally using the line-free normalized continuum region near the absorption line. It is the rms dispersion over one resolution element. For FUV XDL detector, one resolution element consists of six pixels in segment A and ten pixels in segment B (COS instrument handbook; \citealt{Debes16}). }
\label{fig3}
\end{figure*}

To compute $v_{\rm rot}$, we use the \ion{H}{1} 21-cm emission-line observation from the Very Large Array (VLA; \citealt{Gratier10}); the spectral resolution is 1.29$\kms$ and the angular resolution is 25" ($\sim100$ pc at the distance of M33). The \ion{H}{1} 21-cm spectrum at the position of each sightline is shown in the first row of Fig \ref{fig3}.  We fit a single Gaussian function to each spectrum to decide the centroid velocity of the line, which we adopt as $v_{\rm rot}$ (Table \ref{tb1}). Note that we do not attempt to shift the UV absorption-line centroids to line up with the \ion{H}{1} 21-cm peak emission, since it is not clear whether the neutral and the ionized gas associated with M33's ISM are co-spatial and/or kinetically connected. 

In addition, we make use of the \ion{H}{1} 21-cm spectra from the Arecibo Galaxy Environment Survey (AGES; \citealt{Auld06, Keenan15}) that has a spectral resolution of $\sim5\kms$ and angular resolution of $\sim4'$. The AGES data reaches a column density limit of $\sim1.5\times10^{17}$ cm$^{-2}$ over $10\kms$. Its large beam size and high sensitivity helps to map the diffuse gas pervading M33's ISM and its disk-halo interface. We also examine the \ion{H}{1} 21-cm spectra obtained from the GALFA-\ion{H}{1} survey \citep{Peek11} that has the same angular resolution as the AGES data. Similar \ion{H}{1} 21-cm emission line profiles are found. We use the AGES spectra for later analysis since they have better column density sensitivity than GALFA-\ion{H}{1}. The AGES spectra are shown in the second row of Fig \ref{fig3}. We discuss the use of these spectra in Section \ref{sec6.2}. 

\subsection{Hydrogen Column Density in Front of the Stars}
\label{sec2.5}

To assess the metallicity of M33's ISM and the disk-halo interface as discussed in Section \ref{sec6.2} and \ref{sec6.3}, here we calculate the total column density of neutral hydrogen (both in atomic and molecular forms) along the lines of sight in front of our target stars. We cannot pursue this hydrogen measurement with \ion{H}{1} 21-cm emission since it would include all of the \ion{H}{1} both in front of and behind the stars. In addition, the \ion{H}{1} 21-cm spectra were obtained with larger beam size than that of the COS aperture, thus may introduce up to a factor of ten uncertainty in \ion{H}{1} column density due to unresolved sub-structures within the beam (e.g., \citealt{Tumlinson02, Welty12}). On the other hand, estimates based on Ly$\alpha$ $\lambda$1215 absorption line are not reliable since this line is saturated and, distorted due to stellar P Cygni profiles and the dominant Galactic emission (see Fig \ref{fig2}). 

Here we use a method first introduced by \cite{Bohlin78}, which correlates the column density of neutral hydrogen $N_{\rm H}$($=N_{\rm HI}+2N_{\rm H_2}$) in front of the corresponding stars with their color excess E(B-V) -- the gas-to-dust ratio. The relation is established linearly as 
% (\citealt{Bohlin78}; \citealt{Shull85}; \citealt{Diplas94})
\begin{equation}
N_{\rm H}=f\times5.8\times10^{21}\ {\rm E(B-V)} \ {\rm cm}^{-2}\ {\rm mag}^{-1}.
\label{eq1}
\end{equation}
where $f=1.0$ for the Galactic value \citep{Bohlin78}, and $f=2.8\ (5.0)$ for the LMC (SMC) gas-to-dust ratio (\citealt{Welty12, Tumlinson02}). 

Using E(B-V) to infer $N_{\rm H}$ relies on the assumption that local environment does not change dramatically from sightline to sightline. This is reasonable as the gas-to-dust ratio is relatively constant over a galaxy's disk \citep{Sandstrom13} and the dust extinction curve only weakly varies across kpc scales \citep{Schlafly16}. Since the metal content of M33's ISM is not yet well understood, in Table \ref{tb3} we provide log $N_{\rm H}$ values in three cases assuming M33 is MW-, LMC-, or SMC-like. The MW-like log $N_{\rm H}$ has uncertainty of $\lesssim0.2$ dex \citep{Bohlin78}, while the LMC- and SMC-like log $N_{\rm H}$ have rms deviations of 0.22 and 0.34 dex, respectively \citep{Welty12}. We suggest the LMC-like values are likely to be the closest to the M33's since these two galaxies are similar in galaxy mass and metallicity (\citealt{Crockett06, DOnghia15}). The MW-like and SMC-like values serve as the metal-rich and metal-poor guesses, respectively; they provide lower and upper log $N_{\rm H}$ estimates, which are consistent with the quoted LMC-like rms deviation. In Section \ref{sec6.3} and Table \ref{tb5}, we provide metallicity estimates using the LMC-like log $N_{\rm H}$ values. We note that the total \ion{H}{1} column density (integrated from $v_{\rm d}=-30\kms$ to $+30\kms$ for each sightline) is log $N_{\rm HI}\sim21.0$ cm$^{-2}$, consistent with the ones derived from E(B-V). Overall, the LMC-like scenario is an approximation given the current limited knowledge of the actual metal content of M33’s ISM. The reader should interpret our provided N$_{\rm H}$ and metallicity values (Section \ref{sec6.3}) with caution.

%\input t3.tex

%%%%%%%%%%%%%%%%%%%%

\begin{table}
\renewcommand{\arraystretch}{0.9}
\caption{E(B-V) and log $N_{\rm H}$}
% \footnotesize
\begin{center}
\begin{tabular}{X{0.2cm}X{1.0cm}X{1.2cm}X{1.2cm}X{0.8cm}X{0.8cm}X{0.8cm}}
\hline
\hline
ID & (B-V)\tablenotemark{a} & (B-V)$_{0}$\tablenotemark{b} & E(B-V)\tablenotemark{c} & \multicolumn{3}{c}{log $N_{\rm H}$} \\
    & (mag) & (mag) & (mag) & (cm$^{-2}$)   & (cm$^{-2}$)   & (cm$^{-2}$) \\
\hline
    &       &       &       & (MW)          & (LMC)         & (SMC) \\
\hline 
S1 & -0.21 & -0.32 & 0.07	& 20.6 & 21.1 & 21.3 \\
S2 &  0.43 & N/A   & 0.24	& 21.1 & 21.6 & 21.8 \\
S3 & -0.16 & -0.30 & 0.10 	& 20.8 & 21.2 & 21.5 \\
S4 &  0.48 & N/A   & 0.20	& 21.1 & 21.5 & 21.8 \\
S5\tablenotemark{d} &  0.24 & N/A   & 0.20	& 21.1 & 21.5 & 21.8 \\
S6 & -0.14 & -0.30 & 0.12	& 20.8 & 21.3 & 21.5 \\
S7 & -0.13 & -0.32 & 0.15	& 20.9 & 21.4 & 21.6 \\
\hline
\hline
\end{tabular}
\end{center}
\tablenotetext{a}{Observed B-V color including the foreground Galactic extinction. S1, S2, S3, S6, S8: \cite{Massey06}; S7: \cite{Neugent11}; S4, S5: B mag is from \cite{Drissen08} and V mag from \cite{Neugent11}.}
\tablenotetext{b}{Intrinsic color. S3, S6: \cite{Wegner94}. S1, S7: see Section \ref{sec2.5} for the derivation. } 
\tablenotetext{c}{E(B-V) corrected for the foreground Galactic value. }
\tablenotetext{d}{For S5, L06 found log $N_{\rm HI}=20.75\pm0.3$ (log N$_{\rm HI}=21.07^{+0.17}_{-0.24}$) by fitting the damping wing of the Ly$\beta$ (Ly$\alpha$) line. These numbers are consistent with our MW-like value, given that the molecular gas content is negligible along S5 as indicated by L06. See Section \ref{sec2.5} for detailed discussion. }
\label{tb3}
\end{table}

%%%%% Table 3 end %%%

We find the E(B-V) values from a number of sources. For S2, which was named B324 in their paper, \cite{Humphreys13} found a V-band total extinction of $A_{\rm v}=0.9$ mag. Assuming $R_{\rm v}=3.2$ mag as suggested \citep{Humphreys13}, we obtain a total color excess of 0.28 mag, which gives E(B-V)=0.24 mag for S2 after correcting for foreground Galactic extinction ($\sim$0.04 mag; \citealt{Schlegel98, Schruba10}). For S3 and S6, their color excesses are directly calculated using E(B-V)=(B-V)-(B-V)$_0$-0.04 mag. For S4, its E(B-V) is from \cite{Ubeda09} who provided the measurement of E(4405-5495) that is the monochoromatic equivalent of E(B-V) assuming 4405 \AA\ and 5495 \AA\ are the central wavelengths of the B and V filters. For S5, it is from \cite{Bruhweiler03} who measured the E(B-V) of two O-type stars (690A and 690B) in the same COS aperture as S5. 

For S1 and S7, neither a direct measurement of E(B-V) nor an intrinsic color (B-V)$_0$ can be found in the literature. We calculate their E(B-V) as follows. We first infer (B-V)$_0$ for S1 and S7 by searching for WR stars in LMC that have the same spectral type WN9 as S1 and S7. We found two WN9-type stars (BR18 and BR64) with color excess of E(b-v)=0.12 mag and 0.31 mag, respectively (\citealt{Breysacher81, Torres88}). Using E(B-V)=1.21$\times$E(b-v) (\citealt{Turner82, Torres88}), we obtain E(B-V)=0.15 (0.38) mag for star BR18 (BR64). Knowing that the observed color (B-V) of these two stars are -0.16 mag and 0.06 mag respectively, we find the intrinsic color (B-V)$_0$=-0.32 mag for S1 and S7. Then taking into account the observed (B-V) values of S1 and S7, and correcting for foreground Galactic extinction, we find their E(B-V)=0.07 mag and 0.15 mag, respectively. 

We note that the log $N_{\rm HI}$ value along S5 has been determined by Lebouteiller (\citeyear{Lebouteiller06}; henceforth L06) in a study of the neutral ISM of M33. L06 observed Ly$\beta$ (Ly$\alpha$) with the {\it FUSE} ({\it IUE}) instrument, and found log $N_{\rm HI}=20.75\pm0.3$ (log N$_{\rm HI}=21.07^{+0.17}_{-0.24}$) by fitting the damping wing of the saturated Ly$\beta$ (Ly$\alpha$) along S5. These values are similar to our MW-like value for S5 (log $N_{\rm H}=21.1$) derived using the E(B-V) method, but significantly lower than the LMC- and SMC-like numbers, as shown in Table \ref{tb3}. The difference is not likely due to a significant presence of H$_{2}$, since L06 found only a small contribution from molecular gas along S5, log $N_{\rm H_2}=16.86$, from H$_2$ absorption lines in the {\it FUSE} spectra. We suggest that the log $N_{\rm HI}$ estimates from L06 may be lower limits due to: (1) contamination by complicated stellar \ion{O}{6} (\ion{N}{5}) P-Cygni profiles  contaminating the Ly$\beta$ (Ly$\alpha$) damping wings and, (2) saturated Ly$\beta$ (Ly$\alpha$) absorption due to MW's ISM separated from M33's ISM absorption by only $\sim200\kms$. While our MW-like value is consistent with their estimates within the uncertainty, such a comparison may not be overly informative due to the unavoidably large systematic uncertainties inherent in both methods.

%------------------------------ Results -----------------------------
\section{General Ion Properties}
\label{sec3}

% In this section, we discuss the ion properties as inferred from the Voigt-profile fitting. We only consider \ion{Si}{4}, \ion{Fe}{2}, \ion{S}{2}, and \ion{P}{2} since their absorption lines are not saturated. Hereafter we address \ion{Si}{4} as the ``warm ion" and, \ion{S}{2}, \ion{P}{2}, and \ion{Fe}{2} the ``cool ions". We show the continuum-normalized {\it HST}/COS absorptions lines and their Voigt profiles\footnote{Note that absorption lines of \ion{S}{2}, \ion{Fe}{2}, and \ion{Si}{4} are actually multiplets; in Fig \ref{fig3} we only show one line for each ion for clarity since all the lines of the same ion are fit simultaneously (see Section \ref{sec2.3}). A full display of all the lines can be found in Fig \ref{figc7}. } in rows 3-6 of Fig \ref{fig3}. Both the cool ions and warm ion show strong strong absorption from MW's ISM at $v_{\rm d}>150\kms$ ($\vlsr\sim0\kms$), as noted by the dotted lines in Fig \ref{fig3}. The M33-associated absorption lines are at $v_{\rm d}\sim0\kms$ ($\vlsr<-150\kms$). 

In this section, we discuss the ion properties as inferred from the Voigt-profile fitting. In row 3--7 of Fig \ref{fig3}, we show the continuum-normalized absorption lines of \ion{P}{2}, \ion{Fe}{2}, \ion{S}{2}, \ion{Si}{2}, and \ion{Si}{4}\footnote{Note that absorption lines of \ion{S}{2}, \ion{Fe}{2}, \ion{Si}{2}, and \ion{Si}{4} are actually multiplets; in Fig \ref{fig3} we only show one line for each ion for clarity. A full display of all the lines can be found in Fig \ref{figc7}. }. Among these transition lines, \ion{P}{2}, \ion{Fe}{2}, \ion{S}{2}, and \ion{Si}{4} are not saturated, thus they are fitted with Voigt profiles as discussed in Section \ref{sec2.3}. \ion{Si}{2} $\lambda$1304 is not Voigt-profile fitted since this line has saturated core at $v_{\rm d}\sim0\kms$ along each sightline. We do not show other \ion{Si}{2} lines or \ion{C}{2}, \ion{Si}{3} lines in Fig \ref{fig3} because they are all heavily saturated and blended both at the velocity of M33's ISM as well as MW's ISM; therefore, they are not informative in evaluating the ion properties at intermediate velocities as we show in the following. For an inspection of these saturated lines, please see Fig \ref{figa1} and \ref{figc7}.

Since \ion{Si}{4} requires $\sim33.5$eV to ionize from \ion{Si}{3}, while \ion{S}{2}, \ion{P}{2}, \ion{Fe}{2}, and \ion{Si}{2} only need $\sim10$eV to produce from their neutral forms, hereafter we address \ion{Si}{4} as the ``warm ion" and,  \ion{S}{2}, \ion{P}{2}, \ion{Fe}{2}, and \ion{Si}{2} the ``cool ions"\footnote{The ``warm" and ``cool" do not necessarily correspond to the temperature ranges used by other authors (e.g., \citealt{Werk14}); we use the two terms here to reflect the different ionization states of the ions. }. Both the warm and cool ions show strong absorption lines from MW's ISM at $v_{\rm d}\gtrsim150\kms$ ($\vlsr\sim0\kms$), as noted by the dotted lines in Fig \ref{fig3}. The M33-associated absorption lines are at $v_{\rm d}\sim0\kms$ ($\vlsr\lesssim-150\kms$), consistent with the peaks of the corresponding \ion{H}{1} 21-cm emission from M33's ISM (row 1--2). In the following, we summarize the main properties of the M33-associated absorption lines:

\begin{enumerate}

\item On all sightlines, apart from the $v_{\rm d}\sim0\kms$ components associated with M33's ISM, we fit additional velocity component(s) at intermediate velocities to the warm and cool ion lines. We find that the warm ion \ion{Si}{4} consistently shows additional absorption components along each sightline, with $v_{\rm d}$ ranging from $+39\kms$ to $+106\kms$ (indicated by ``*" sign in Table \ref{tb2}). The mean value is $v_{\rm d}=+$67$\kms$. Once corrected for inclination, this corresponds to a warm-ionized inflow toward M33 at $+$67/cos(56$^{\degree}$)$\sim+120\kms$ (shown by the dotted line in the left panel of Fig \ref{fig4}; see point 7). In Fig \ref{fig3}, we highlight these \ion{Si}{4} components with intermediate velocities in blue. We also find absorption features at similar intermediate velocities in cool ions, such as those ions along S6 and \ion{Fe}{2} along S5, although they do not consistently exist among all the sightlines. Since these intermediate-velocity absorption components have mean $v_{\rm d}$($=+67\kms$) noticeably separated from the M33's ISM absorption ($v_{\rm d}\sim0\kms$), hereafter we call these features the ``non-disk" components. And we call the absorption components at $v_{d}\sim0\kms$ the ``disk" components. {\it The consistently present non-disk components in \ion{Si}{4} suggests that there exists a disk-wide, warm-ionized gas accreting toward M33.} This accreting flow is most likely multi-phase, as the non-disk components can also be observed in the cool ions (\ion{Fe}{2}, \ion{S}{2}, and \ion{P}{2}) along some sightlines, and may as well be commonly present in saturated \ion{C}{2}, \ion{Si}{2}, and \ion{Si}{3} lines (also see Point 3).

\item The AGES \ion{H}{1} 21-cm spectra in row 2 show extended \ion{H}{1} wings at the intermediate velocities where the non-disk \ion{Si}{4} are detected, also supporting the multi-phase scenario of the accreting gas. The VLA spectra on the top row indicate there may be some \ion{H}{1} emission within the highlighted ranges; however, after fitting baselines and checking the sensitivity of the VLA data, we find there is no signal above the noise level given by \cite{Gratier10}. 

\item The saturated ions (i.e., \ion{Si}{2}, \ion{Si}{3}, and \ion{C}{2}) are consistent with the accreting flow being multi-phase. In row 6, we show the continuum-normalized profiles of \ion{Si}{2} $\lambda$1304 (night-only data). We do not Voigt-profile fit this line as the core is saturated. Along some sightlines (e.g., S1, S3, S6)
where the non-disk velocity ranges (blue shades) fall out of the core, we find significant \ion{Si}{2} absorption, consistent with the non-disk \ion{Si}{4} components.

\item The line profiles and line widths of the non-disk \ion{Si}{4} components vary from sightline to sightline. Such discrepancy is unlikely due to contamination by stellar winds since those stellar features are much broader and have been fitted as continua (see Section \ref{sec2.3}). In addition, the continuum-normalized \ion{Si}{4} $\lambda\lambda$1393, 1402 show similar line profiles despite their different wavelengths (see Fig \ref{figc7}), which also suggests that local stellar features are unlikely to cause the variation seen in the data.

\begin{figure}[t!]
\centering
\includegraphics[width=0.47\textwidth]{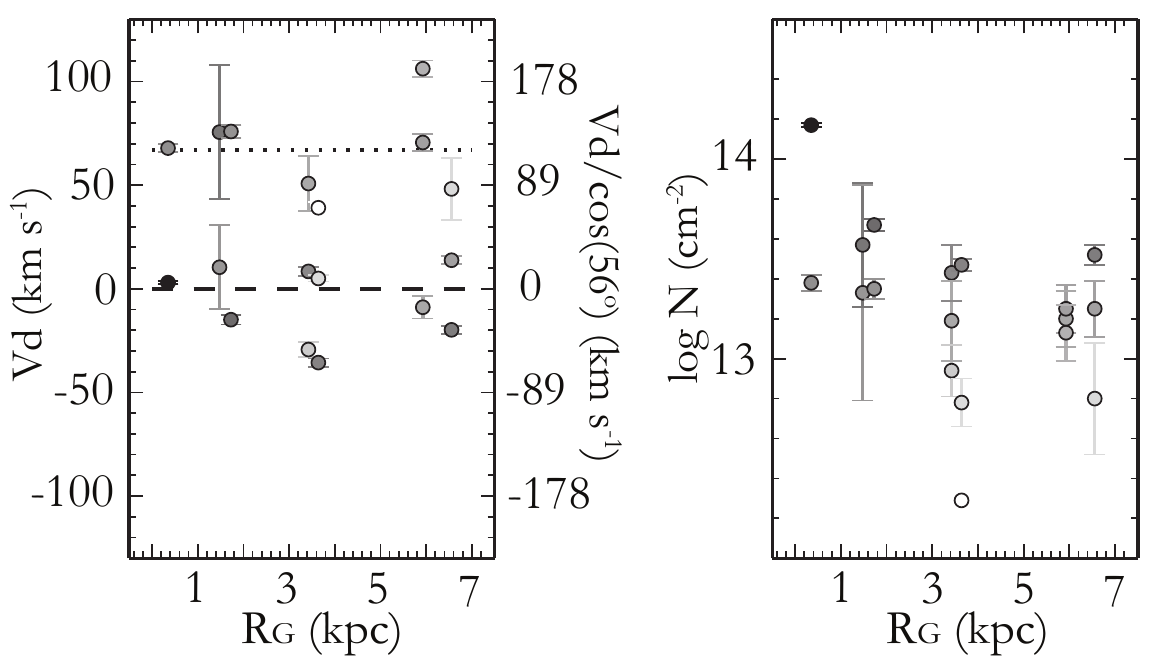}
\caption{Radial distribution of $v_{\rm d}$ (left) and log $N$ (right) for \ion{Si}{4}. Data points are color-coded in gray scale such that darker color means larger log $N$. The $R_{\rm G}$ on the x-axis shows the inclination-corrected galactocentric distance. In the left panel, the left y-axis indicates the observed/projected velocity for each absorption component $v_{\rm d}$ while the right y-axis shows the inclination-corrected values $v_{\rm d}$/cos(56$^{\degree}$). The dotted line shows the mean velocity $v_{\rm d}$=+67$\kms$ (or +67/cos(56$^{\degree}$) $\sim$ +120$\kms$ if corrected for inclination) of the non-disk \ion{Si}{4} components. }
\label{fig4}
\end{figure}

\item S6 shows two non-disk \ion{Si}{4} components. With limited S/N ratio (=6), it is less certain whether the line has one broad component or two narrower ones but, in general, both components are consistent with the non-disk components seen along other sightlines. S6 is of particular interest since within the \ion{Si}{4} non-disk velocity range, its \ion{H}{1} 21-cm spectrum shows the least contamination from the dominant disk emission. We make use of this sightline to calculate the \ion{H}{1} column density of the non-disk component in Section \ref{sec6.2}. 

% \item For the cool ions, \ion{Fe}{2}, \ion{P}{2}, \ion{S}{2}, and \ion{Si}{2}, also present intermediate-velocity components along a couple of sightlines. Since these ions are in a weakly ionized phase that is different from \ion{Si}{4} and their kinematics does not show consistency from sightline to sightline, we do not intend to analyze these ions in this work. We note  Likely, these cool ions are associated with M33's ISM since their absorption profiles are mostly centered at $v_{\rm d}\sim0\kms$. 

\item In Table \ref{tb2}, we show the Voigt-profile fitted ($v_{\rm d}$, $b$, log $N$) for each sightline. For the warm ion \ion{Si}{4}, the non-disk component has mean $\langle N_{\rm SiIV}\rangle=10^{13.24\pm0.21}$ cm$^{-2}$, and the disk component has $\langle N_{\rm SiIV}\rangle=10^{13.55\pm0.08}$ cm$^{-2}$. As a comparison, we find that MW IVCs have $\langle N_{\rm SiIV}\rangle=10^{13.17\pm0.07}$ cm$^{-2}$ \citep{Shull09}, which suggests the non-disk \ion{Si}{4} components of M33 are consistent with IVC-like clouds. For the cool ions,  we find mean column densities of $\langle N_{\rm SII}\rangle=10^{15.07\pm0.04}$ cm$^{-2}$, $\langle N_{\rm FeII}\rangle =10^{14.34\pm0.04}$ cm$^{-2}$ and $\langle N_{\rm PII}\rangle = 10^{13.55\pm0.06}$ cm$^{-2}$, respectively, taking into account both the disk and non-disk components. The Voigt-profile fits of these cool ions show considerably large $b$ values along a few sightlines, which might result from the COS spectral resolution that fails to resolve underlying narrower components. 

\item Specifically, we show the radial distribution of $v_{\rm d}$ and log $N$ of the warm ion \ion{Si}{4} in Fig \ref{fig4}. Each data point is color-coded in gray scale according to its log $N$: darker color means higher log $N$. For example, at $R_{\rm G}\sim$ 0.5 kpc, a \ion{Si}{4} component with $v_{\rm d}\sim$ 0$\kms$ in the left panel can be found with log $N\sim14.2$ cm$^{-2}$ in the right. Clearly \ion{Si}{4} shows a significant asymmetric distribution of $v_{\rm d}$ toward positive (inflow) velocities. Along each sightline, the disk component generally has larger column density than the non-disk component. We also examine the radial distribution $v_{\rm d}$ and log $N$ for the cool ions \ion{S}{2}, \ion{P}{2}, and \ion{Fe}{2}. Their centroid velocities $v_{\rm d}$ mostly concentrate near $v_{\rm d}\sim0\kms$, and there is no obvious radial trend in column density.
\end{enumerate}

Apart from the disk-wide accreting flow detected in the warm ion \ion{Si}{4}, our sightlines show some low-velocity, blue-shifted components with $-40\lesssim v_{\rm d}\lesssim0\kms$ in both warm and cool ions, indicating a tentative detection of slow winds or disk turbulence in M33; we discuss this in Section \ref{sec7.3.1}. In addition, we find a very high-velocity absorption feature with $-200\lesssim v_{\rm d} \lesssim -100 \kms$ (or $\vlsr\sim-350\kms$) in \ion{C}{2}, \ion{Si}{2}, and \ion{Si}{3}. This could potentially be the outflows from M33, however, the scenario is complicated by other origin possibilities as discussed in Section \ref{sec7.3.2} and Appendix \ref{appA}.

We note that the same COS spectrum of S1 was analyzed by Welsh \& Lallement (\citeyear{Welsh13}; henceforth WL13), finding multiple narrow components in cool ions for which we are only be able to fit broad Voigt-profiles. In their work, the multiple narrow-component fits were achieved by restricting five components at fixed velocities (varying by only $\pm$10$\kms$) as indicated from \ion{Ca}{2} lines \citep{Welsh09}. Since it is unclear whether the \ion{Ca}{2}-bearing gas and the low-ion-bearing gas are kinematically similar, this fixed-velocity fitting procedure is likely to lead to artificial components. Therefore, we do not compare our component fits to the results shown in WL13. In addition, we note that their continuum fittings for \ion{O}{1} $\lambda$1302 and \ion{Si}{2} $\lambda$1304 are inaccurate due to the omission of recognizing the air-glow emissions at nearby location. We provide in Fig \ref{figa1} the night-only spectra of \ion{Si}{2} $\lambda$1304, which are weaker than those shown in WL13.

\begin{figure*}[t!]
\includegraphics[width=\textwidth]{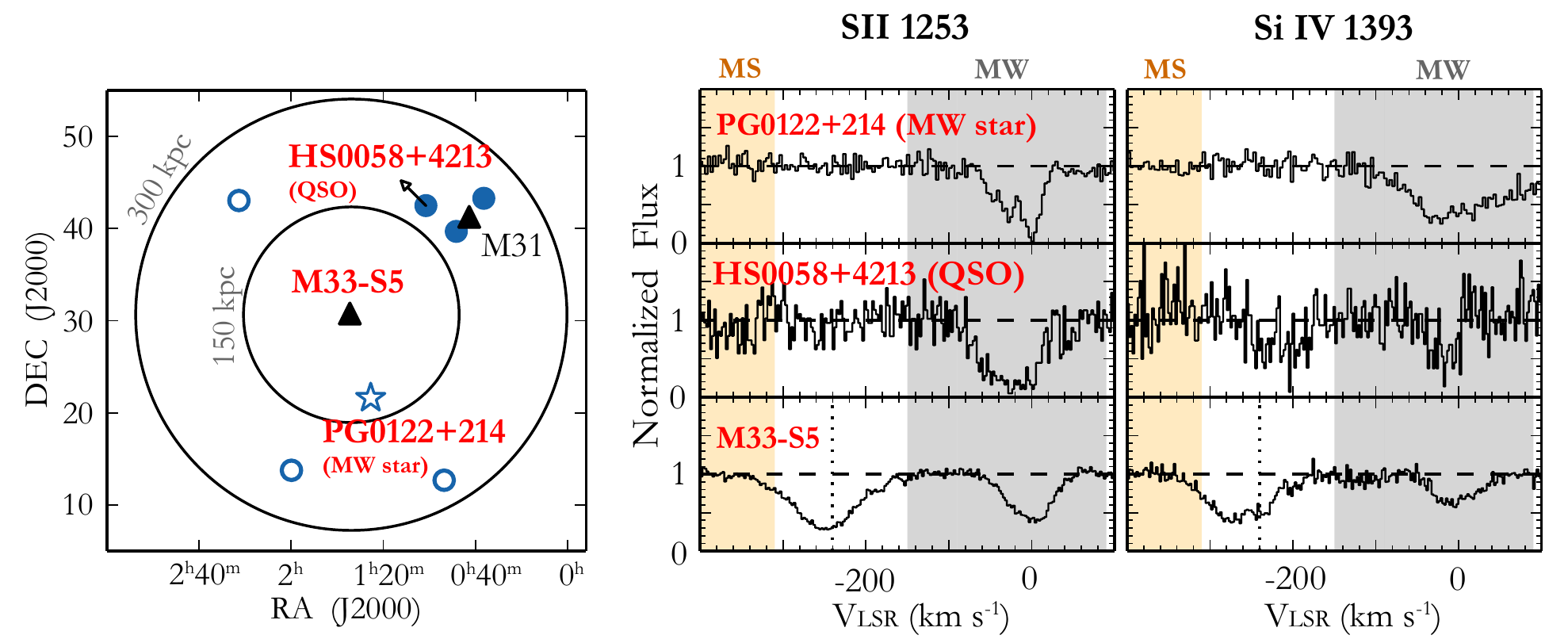}
\caption{Left: The positions of the MW halo star PG0122+214 (open star), QSO sightlines (open and solid circles) within $\sim300$ kpc of M33, the M31 disk (triangle), and S5 toward M33 are shown. The M33 sightlines S1--S7 are at almost the same spot as S5. For QSO sightlines, solid (open) circles indicate detection (non-detection) of \ion{Si}{4} at $\vlsr<-150\kms$; we show one of the QSOs (HS0058+4213) spectra in the middle row of the right panel. Right: Normalized \ion{S}{2} $\lambda$1253 and \ion{Si}{4} $\lambda$1393 absorption lines of the MW halo star PG0122+214 (top), of the QSO HS0058+4213 (middle), and of M33-S5 (bottom). Note that the velocities of these spectra are in the LSR frame. Our M33 absorption lines appear at $\vlsr\lesssim-180\kms$ while the MW's absorption lines are at $\vlsr\geq-90\kms$. The dotted lines in the bottom indicate the systemic velocity ($\vlsr=-240.3\kms$) of M33's gas disk at the position of sightline S5 (Table \ref{tb1}). The highlighted light yellow and gray shades are the ``MS-related" and ``MW-related" components as defined in \cite{Lehner15}. } 
\label{fig5}
\end{figure*}

%=======================================================================
\section{Association of The Non-Disk Components with M33}
\label{sec4}

As we show in Section \ref{sec3}, we find consistently redshifted non-disk components in \ion{Si}{4}, suggesting that there exists a disk-wide, warm-ionized accreting flow toward M33. The accreting flow is most likely multi-phase as non-disk components can also be seen in cool ions (\ion{Fe}{2}, \ion{S}{2}, and \ion{P}{2}) along some sightlines and may as well commonly exist in saturated \ion{C}{2}, \ion{Si}{2}, and \ion{Si}{3} lines. The non-disk components follow the overall rotation of M33 and are almost certainly associated with this galaxy; however, given the Local Group spatial and kinematic environment, other sources should be considered. Apart from the M33 disk-halo accreting layer and M33 halo cloud scenarios modeled in Section \ref{sec5}, here we focus on the possibilities of ion absorbers from MW's ISM, MW HVCs, the Magellanic Stream, or M31's CGM.

First, MW's ISM and the Magellanic Stream are unlikely to have created the absorption lines. We show in Fig \ref{fig5} that, in the LSR frame, MW's ISM is at only a few tens of $\kms$ and the Magellanic Stream is at -310 to -450 $\kms$ in directions near M33 as shown in Fig \ref{fig5} (\citealt{Fox14, Lehner15}). Our M33 absorption lines fall in between these two velocity ranges.

Second, absorbers from MW HVCs are unlikely intercepting our sightlines toward M33 at the intermediate velocities. We compare our spectra with those from a nearby MW halo star, PG0122+214, which is at a distance $\sim10$ kpc from us. In the right panel of Fig \ref{fig5}, we display the \ion{S}{2} and \ion{Si}{4} lines\footnote{We use \ion{S}{2} $\lambda$1253 and \ion{Si}{4} $\lambda$1393 because 1. they generally have high S/N than other ion lines, and 2. they have higher oscillator strengths among their available multiplets. } from PG0122+214 in the top row, and those from sightline S5 toward M33 in the bottom row; clearly the MW halo star does not have \ion{S}{2} and \ion{Si}{4} absorption lines at $\vlsr<-150\kms$ where M33-S5 shows strong absorption. Since the angular separation between PG0122+214 and M33-S5 is only $\sim9^{\degree}$, if there exists an HVC within 10 kpc of the MW halo that covers all our M33 sightlines but not PG0122+214, the HVC's physical size should only be $\lesssim1.6$ kpc. This is contrary to current knowledge of classic MW HVCs, which finds them to be large \ion{H}{1} complexes \citep{Wakker91c} with extended ionized gas (e.g., \citealt{Lehner11b, Fox04}). Thus, within 10 kpc there are no MW HVCs with $\vlsr<-150\kms$ in direction toward M33. In addition, MW HVCs beyond 10 kpc are unlikely as we do not find relevant strong absorption from nearby QSOs (shown as open circles in Fig \ref{fig5}).

Third, we investigate the situation that the non-disk warm and cool absorbers are from M31's CGM. We examine the QSO sightlines that are within 300 kpc of M33. These sightlines were studied by \cite{Lehner15}, who found three QSO sightlines (solid circle in Fig \ref{fig5}) with M31-related \ion{Si}{4} absorption within 50 kpc of M31's disk,  while others beyond this radius (open circles) do not have significant detection. In the middle row of the right panel, we plot the \ion{S}{2} and \ion{Si}{4} lines from HS0058+4213 -- one of the three QSOs with M31-related \ion{Si}{4} detection. The \ion{S}{2} line does not show absorption at M33's velocity while the \ion{Si}{4} does indicate some overlap at $\vlsr\sim-240\kms$. The \ion{Si}{4} column density of the overlapped portion along this QSO sightline is $\sim10^{13.1}$ cm$^{-2}$ \citep{Lehner15}, lower than the mean non-disk \ion{Si}{4} value ($10^{13.24}$ cm$^{-2}$) that we have found toward M33. Given that our M33 sightlines are $\gtrsim200$ kpc from M31, we expect the \ion{Si}{4} column density of M31's CGM should be lower than $\sim10^{13.1}$ cm$^{-2}$ at M33's position, as inferred from other CGM studies (e.g., \citealt{Werk13}). Our strong detection of the non-disk \ion{Si}{4} toward M33 suggests that the M31 origin is highly unlikely. In addition, the fact that our \ion{Si}{4} lines clearly follow the rotation of M33 disk as modelled in Section \ref{sec5} also state their association with M33.

To conclude, we find that the non-disk warm and cool absorption lines we have observed are unlikely to be associated with MW's ISM, MW HVCs, the Magellanic Stream, or the M31 halo. In the next Section, we perform kinematic modeling and study if the ions could be related to the M33 disk-halo interface or a gas cloud in the M33 halo.

\begin{figure*}[t!]
\includegraphics[width=\linewidth]{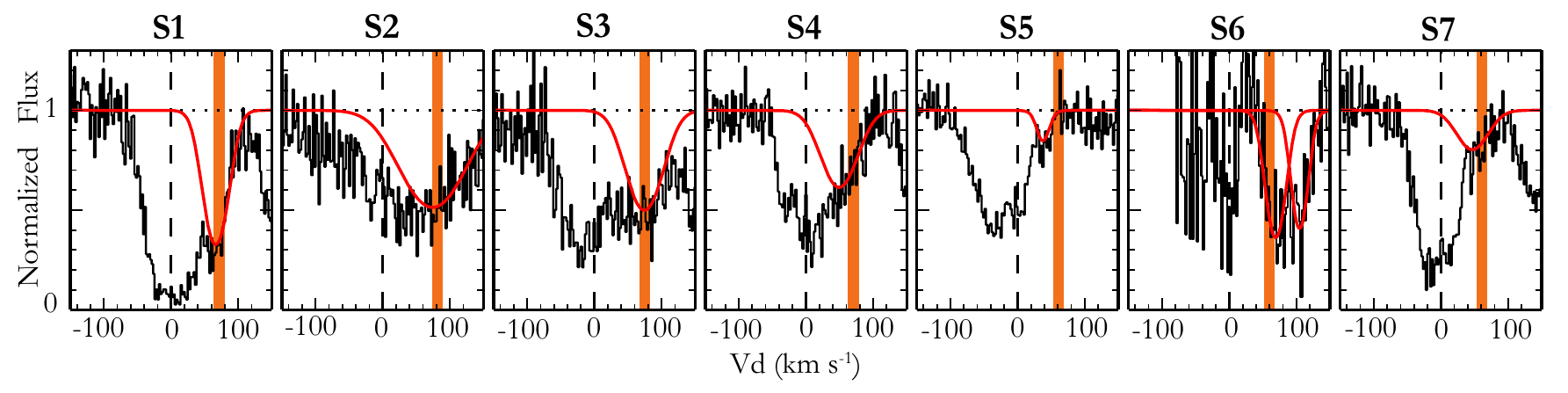}
\caption{The Voigt-profile fitted non-disk \ion{Si}{4} components (red curves) and the model-predicted centroid velocities (vertical lines) from the Accreting Layer model. The \ion{Si}{4} spectra and the fitted Voigt-profiles are the same as those in Fig \ref{fig3}.}
\label{fig6}
\end{figure*}

%=======================================================================

\section{Kinematic Modeling}
\label{sec5}

In this section, we model the kinematics of the accreting flows onto M33 as observed in warm and cool ions. Since the red-shifted, non-disk components are consistently found in \ion{Si}{4}, they provide a complete picture of the flows across the disk and we hereafter focus our analysis on \ion{Si}{4} lines. We consider two scenarios: 1) an accreting layer on top of M33's gas disk, representing the disk-halo interface, and 2) a cloud in the halo of M33 that intercepts all of our sightlines. Hereafter, we refer to these two models as the ``Accreting Layer" model and the ``M33 Halo Cloud" model.

\subsection{The Accreting Layer Model}
\label{sec5.1}

The Accreting Layer model is a layer of warm ionized gas at the disk-halo interface of M33 moving toward the disk. It starts with an \ion{H}{1} tilted-ring model \citep{Corbelli97}, simulating the \ion{H}{1} disk of M33 as a set of concentric tilted rings with each ring gradually differing in inclination, position angle and rotation velocity. The model recovers the velocity field of the \ion{H}{1} disk. We then lift the whole set of tilted rings to some height above the disk to represent the accreting layer. 

Using polar coordinates that originate from the disk center, the line of sight velocity $v(r, \theta)$ of a parcel of ionized gas at position $(r, \theta)$ in the accreting layer can be written as, 
\begin{equation}
\begin{split}
v(r, \theta) & = v(r) cos\theta sin \phi(r) + v_{\rm acc} cos \phi(r) \\
v(r)         & = (v_{\infty}+v_{\rm lag} z_{\rm acc}) tanh(r/ {\Delta}_v)
\end{split}
\end{equation}
where $v_{\infty}$ is the rotation velocity of the gas disk at infinity, $\Delta_v$ is the difference of the rotational velocity between adjacent tilted rings, and $\phi(r)$ is the inclination of each tilted ring. We adopt $\phi$($r$)$\approx \phi_{\rm M33}=$56$\degree$, since $\phi(r)$ varies very little within the star-forming disk. We refer the reader to the work by \cite{Corbelli97} for a detailed discussion of each of these parameters. 

The Accreting Layer model differs from the tilted ring model in that it includes three additional parameters, $v_{\rm acc}$, $z_{\rm acc}$, and $v_{\rm lag}$. $v_{\rm acc}$ is the inflow velocity of the accreting layer perpendicular to the gas disk. We do not model $v_{\rm acc}$ faster than $\sim180\kms$ as beyond this velocity the absorbers would be obscured by absorption from MW's ISM. Note that we use a positive sign for the accretion velocities in our modeling. $z_{\rm acc}$ is the height of the layer above the disk. $v_{\rm lag}$ is the drop-off of the rotation velocity above the disk plane (halo lagging). Here, we model the accreting layer with $v_{\rm lag}=0\kms\ {\rm kpc^{-1}}$ (non-halo-lagging) and, with $v_{\rm lag}=-15\kms\ {\rm kpc^{-1}}$ (halo-lagging) which is a typical value found from observations of the MW and external edge-on galaxies, as well as from numerical simulations (\citealt{Levine08, Marasco11, Marasco12}).  

\begin{figure}[t!]
\includegraphics[width=.5\textwidth, height=0.85\textheight]{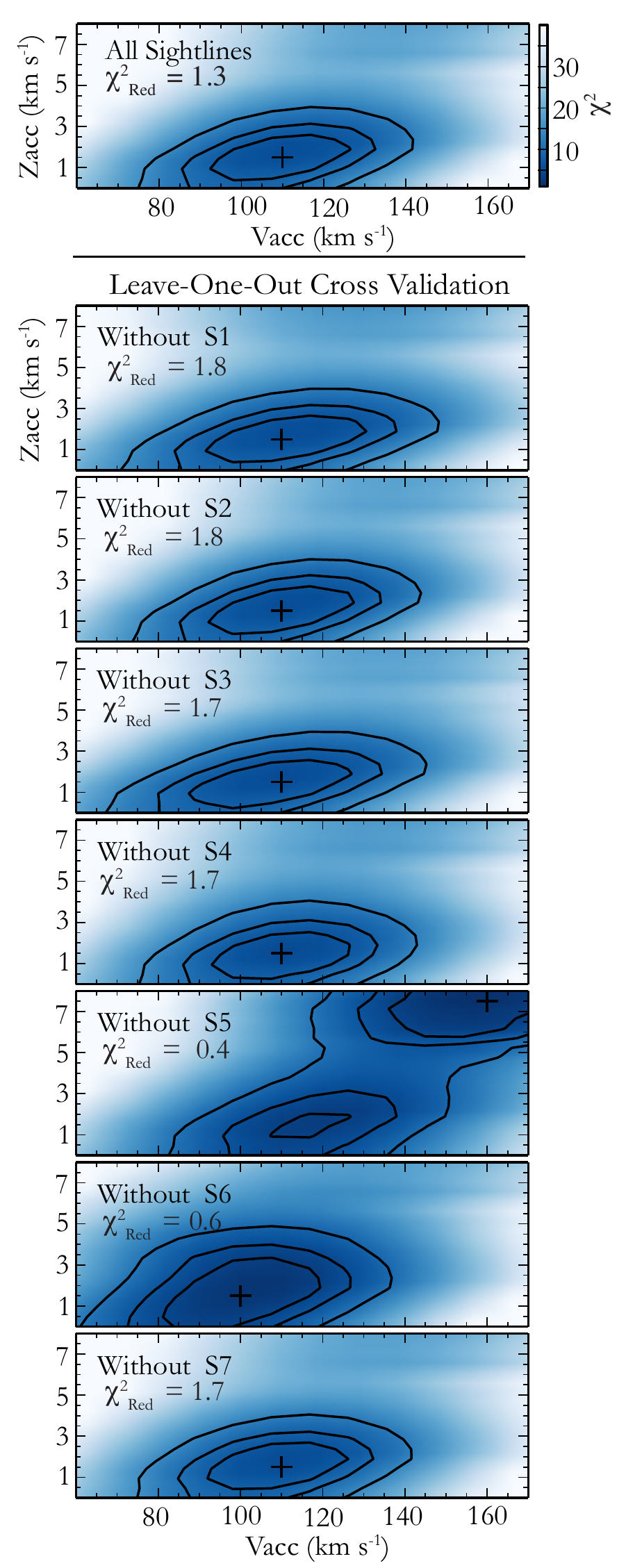}
\caption{The top panel shows the $\chi^2$ distribution of the non-halo-lagging case of the Accreting Layer model using all the sightlines. The minimum $\chi^2$ value is indicated by a cross, and its $\chi^2_{\rm red}$ value is shown in the top left corner. Contours represent 1/2/3 $\sigma$ (68\%/95\%/99\%) confidence levels. The bottom seven panels show the LOOCV of the model as described in Section \ref{sec5.1}; color-coding and labeling are the same as the top panel. }
\label{fig7}
\end{figure}

The model predicts line-of-sight velocity $v_{\rm l,\ i}$ of the layer at the position of each sightline with $i$ representing the sightline ID 1--7. We compare $v_{\rm l,\ i}$ with the observed centroid velocity $v_{\rm d,\ i}$ of the non-disk \ion{Si}{4} components along each sightline, and use $\chi^2$ minimization to find the best fit parameters ($v_{\rm acc}$, $z_{\rm acc}$, $v_{\rm lag}$). The $\chi^2$ is computed as, 
\begin{equation}
\chi^2= \sum_{\rm i=1}^{\rm M=7}\frac{(v_{\rm l,\ i}-v_{\rm d,\ i})^2}{\sigma_{\rm cos}^2+\sigma_{\rm d,\ i}^2}
\label{chi2}
\end{equation}
where $\sigma_{\rm cos}=15\kms$ is the COS velocity calibration uncertainty, and $\sigma_{\rm d,\ i}$ is the uncertainty of velocity centroid from the Voigt-profile fitting\footnote{S6 has two non-disk components: one is at $v_1=70.6\pm4.0\kms$, and the other at $v_2=106.2\pm3.9\kms$. For the kinematic model fitting, we use a mean value $\langle v \rangle=88.4\kms$, and calculate its 1$\sigma$ value through error propagation $\sigma_{\rm d,\ 6}=\frac{1}{2}\sqrt{3.9^2+4.0^2}\kms$}. We find that for the non-halo-lagging case ($v_{\rm lag}=0\kms\ {\rm kpc}^{-1}$), the minimum $\chi^2$ is reached when $v_{\rm acc}=110\kms$, $z_{\rm acc}=$1.5 kpc; its reduced $\chi^2$ value is $\chi^2_{\rm red}=$1.3, indicating a good fit. The halo-lagging case ($v_{\rm lag}=-15\kms\ {\rm kpc^{-1}}$) finds similar results, suggesting our models are not sensitive to velocity drop-off with z-height at the disk-halo interface of M33. Hereafter we only discuss the non-halo-lagging case. 

In Fig \ref{fig6}, we show the performance of the best fit model that has $v_{\rm acc}=110\kms$ and $z_{\rm acc}=1.5$ kpc. The figure visualizes the model-predicted velocity $v_{\rm l,\ i}$ of each sightline in comparison with the respective Voigt-profile fitted non-disk component of \ion{Si}{4}. The model shows a good match with the observed spectra, except that along S6 the model favors the lower velocity component and along S5 the model indicates an absorption component with velocity $\sim20\kms$ higher than that of the observed.  

The top panel in Fig \ref{fig7} shows the distribution of $\chi^2$ as a function of $v_{\rm acc}$ and $z_{\rm acc}$ for the non-halo-lagging case. Contours indicate the confidence limits (P$=0.68/0.95/0.99$) for the best fit model with $v_{\rm acc}=110\kms$ and $z_{\rm acc}=$1.5 kpc. The probability P that the minimum $\chi^2_{\rm red}$ has a certain increment $\delta_{\alpha}$ can be expressed as
\begin{equation}
{\rm Probability}(\chi^2_{\rm red}-\chi^2_{\rm red, min}\leq \delta)\equiv {\rm P}
\end{equation}
. At confidence limits of P$=0.68$ (1$\sigma$), 0.95 (2$\sigma$) and 0.99 ($3\sigma$), the increment $\delta$ is 2.30, 4.61, 9.21 for a model with 2 parameters \citep{Avni76}. The top panel in Fig \ref{fig7} shows that at 1 $\sigma$ confidence level, the best fit parameters are $v_{\rm acc}=110_{-20}^{+15}\kms$ and $z_{\rm acc}=1.5_{-1.0}^{+1.0}$ kpc.

We use the Leave-One-Out Cross Validation (LOOCV) to evaluate the robustness of our model fitting. The method is performed by training the model parameters ($v_{\rm acc}$, $z_{\rm acc}$) using all the sightlines except for one that will be treated as a validation set. When the best fit parameters are found, a prediction is made for the sightline in the validation set; a good set of parameters will predict a velocity similar to the observed one of the validating sightline. In general, this method tests whether the best fit parameters are sensitive to certain sightline(s). We perform LOOCV seven times, each with six sightlines in the training set and the remaining one in the validation set. Every time when a $\chi^2_{\rm red, min}$ value is found at certain ($v_{\rm acc}$, $z_{\rm acc}$), we check if these parameters predict a velocity that matches the observed velocity of the validating sightline. 

\begin{figure*}[t!]
\includegraphics[width=\textwidth]{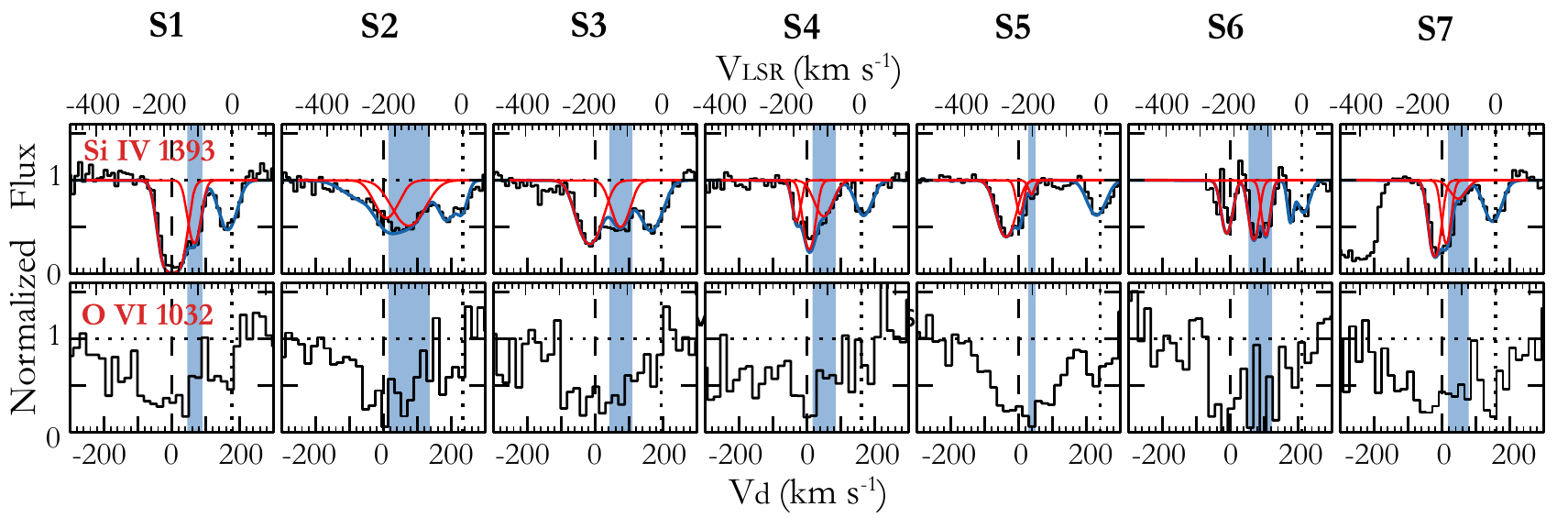}
\caption{The \ion{Si}{4} spectra from {\it HST}/COS (top) compared to the \ion{O}{6} lines observed by {\it FUSE} (bottom). The absorption lines from M33's ISM with $v_{\rm d}=0\kms$ are shown by dashed lines, while those from MW's ISM with $\vlsr=0\kms$ are indicated by dotted lines. The \ion{O}{6} spectra have been binned by five pixels to increase the S/N. The \ion{Si}{4} spectra and the fitted Voigt-profiles are the same as those in Fig \ref{fig3}, but binned by five pixels here to keep the consistency with the {\it FUSE} \ion{O}{6} spectra. The blue patch in each panel highlights the non-disk \ion{Si}{4} components. }
\label{fig8}
\end{figure*}

The result of LOOCV is shown in panels 2--8 of Fig \ref{fig7}. It is shown that when S1/S2/S3/S4/S6/S7 are individually used as the validation set, the model prediction and its $\chi^2$ distribution are very similar to the result ran with all the sightlines (the top panel), and for each experiment the predicted velocity and the observed of the validation sightline show a good match. However, when adopting S5 as the validation set (panel 5), the $\chi^2$ distribution changes dramatically. It predicts two possible minimum $\chi^2$ values: one is similar to that in other panels while the other indicates some fast-moving medium at higher z height. We find that the former can reproduce the centroid velocity as observed in S5, while the latter one fails. This indicates that S5, and its clear non-detection of \ion{Si}{4} above $v_{\rm d}\sim$50$\kms$, is critical for our model fitting in distinguishing different scenarios and it justifies that the original best-fit parameters are robust.

To conclude, we find the observed centroid velocities of the non-disk \ion{Si}{4} components can be well reproduced by assuming an accreting layer at the disk-halo interface of M33. The Accreting Layer model finds the best fit parameters at $v_{\rm acc}=110_{-20}^{+15}\kms$ and $z_{\rm acc}=1.5_{-1.0}^{+1.0}$ kpc.

\subsection{The M33 Halo Cloud Model}
\label{sec5.2}
Alternatively, we consider an M33 halo cloud may exist and cause the non-disk absorption features. We assume the cloud is large enough to cover all the sightlines and, there is no velocity variation across the cloud. If the cloud is at a distance of 10 kpc from the M33 disk, its size should be $\sim$7.5 kpc; if it is 100 kpc away in M33's halo, its size reduces to $\sim$6.5 kpc.   

We vary the velocity of the halo cloud toward M33 ($v_{\rm hc}$) and calculate the model-predicted absorption velocity at the position of each sightline. Since the cloud is far from the M33 disk, its velocity is not influenced by the disk rotation; therefore, the potential absorption lines along different sightlines created by this halo cloud should have the same velocities in the LSR frame. We derive the $\chi^2$ value between the predicted and the observed velocities using Eq \ref{chi2} and study the overall $\chi^2$ distribution. The best fit model is found at $v_{\rm hc}=45_{-5}^{+10}\kms$ with a minimum $\chi^2$ of 27.07 and the corresponding $\chi^2_{\rm red, min}$ of 5.41. This indicates an M33 halo cloud with a uniform velocity does not yield a good fit to the non-disk \ion{Si}{4} component absorption lines. This is because the velocities of our detected non-disk \ion{Si}{4} components consistently indicate the influences of the M33 rotation, a disk-wide, different velocity pattern that an M33 cloud in the distant halo with only uniform velocity would not be able to reproduce.

With our limited number of sightlines, we are unable to place further constraints on whether a non-uniform cloud would cause the observed absorption lines. Different velocity fields across the cloud surface and potentially arbitrary turbulent velocities may cause the variation of centroid velocity seen in the non-disk \ion{Si}{4} components. In addition, future models that incorporate the clumpiness of the halo cloud would be helpful since density variation has been intentionally excluded in our kinematic modeling. 

%===========================================================================
\section{Ionization Condition and Gas-Phase Abundances}
\label{sec6}

In this section, we investigate the ionization sources and the gas metallicity of the accreting flow. Specifically, we study the properties of the non-disk \ion{Si}{4} components since they are consistently found among all the sightlines and their accreting velocities are well reproduced by the Accreting Layer model. In Section \ref{sec6.1} we assess the role of shocks and collisional ionization in creating the non-disk \ion{Si}{4} components. In Sections \ref{sec6.2} and \ref{sec6.3} we put constraints on the metallicity of the non-disk gas at the disk-halo interface and the ISM of M33, respectively. This helps to determine whether the disk-halo gas has been metal-enriched. 

\subsection{The Ionization Mechanism of \ion{Si}{4}}
\label{sec6.1}

The accreting layer discussed in the previous section well reproduces the observed velocities of the non-disk \ion{Si}{4} components. In this model, shocks could be induced as the disk-halo gas accretes at a velocity of $v_{\rm acc}=110\kms$ onto the disk and, \cite{Allen08} showed that slow radiative shocks moving at $v_{\rm s}\lesssim$170$\kms$ would produce collisionally ionized gas behind the shocks (see also \citealt{Dopita96}). This is of particular interest as we find that the \ion{Si}{4} absorption-line profiles are similar to those of the \ion{O}{6} despite the difference of aperture size of {\it HST}/COS and {\it FUSE} data (see Fig \ref{fig8}). The similarity between the profiles may suggest some physical connection. Since \ion{Si}{4} requires 34 eV to produce while \ion{O}{6} needs 114 eV, it is unlikely both ions are created under the same ionization conditions; however, it is worth considering whether these two ions are created by collisional ionization behind shocked gas as the accreting layer plummets toward the disk.

In their models, \cite{Allen08} predicted ion column densities as a function of shock velocity, gas metallicity, gas density, and magnetic field strength. Their models show that for $v_{\rm s}\lesssim170\kms$, the \ion{O}{6} column density is $\lesssim10^{14.0}$ cm$^{-2}$ while the \ion{Si}{4} column density is $\lesssim10^{12.30}$ cm$^{-2}$. As a comparison, we calculate the \ion{Si}{4} and \ion{O}{6} column densities of the non-disk gas which are highlighted in blue in Fig \ref{fig8}. For \ion{Si}{4}, the column densities can be found in Table \ref{tb2}; for \ion{O}{6}, the column densities are integrated over the highlighted velocity ranges using the AOD method\footnote{The quality of the {\it FUSE} spectra does not allow for a reliable Voigt-profile decomposition of the \ion{O}{6} line profiles. } (see Appendix \ref{appA}; \citealt{Savage96}). The mean values are $\langle N_{\rm SiIV}\rangle=10^{13.24}$ cm$^{-2}$ and $\langle N_{\rm OVI}\rangle=10^{14.16}$ cm$^{-2}$, both larger than those predicted by shock models, suggesting that shocks are not the dominant mechanism in creating these two ions.

From a broader point of view, we investigate whether general collisional ionization processes could be the dominant producer of \ion{Si}{4} at the disk-halo interface. It has been suggested that if an ion is collisionally ionized in radiatively cooling flows, its total column density is proportional to the Doppler width $b$ of the corresponding absorption lines (e.g., \citealt{Heckman02, Lehner11a}). Following \cite{Heckman02} and \cite{Lehner11a}, the column density of an ion $X_{\rm i}$ in a cooling flow is $N_{\rm X_{i}}$=$v_{\rm cool}t_{\rm cool}n_{\rm cool}$, where $v_{\rm cool}$ is the velocity, $t_{\rm cool}$ the cooling time, and $n_{\rm cool}$ the number density. Assuming the ideal gas law and introducing the cooling function $\Lambda$ to solve for $t_{\rm cool}$ and $n_{\rm cool}$, one would find $N_{\rm X_{i}} = 4.34\times(3/2+s)k_{\rm B} T v_{\rm cool} \frac{Z f_{\rm i}}{\Lambda}$, where $s$ equals to 0 (1) for isochoric (isobaric) cooling, $f_{\rm i}$ the gas fraction of the corresponding ion, and Z the metallicity. \cite{Heckman02} suggested that $v_{\rm cool}$ could be identified as the non-thermal broadening $b_{\rm nth}$ such that the Doppler width is $b^2=b_{\rm th}^2+b_{\rm nth}^2=2k_{\rm B}T/Am_{\rm p}+v_{\rm cool}^2$, where $A$ is the mass number and $m_{\rm p}$ the proton mass. Combining the above equations to solve for $N_{\rm X_i}$ and $b$, an analytical relation can be laid out as
\begin{equation}
N_{\rm X_{i}} = 4.34\times(3/2+s)k_{\rm B} T \sqrt{b^2 - \frac{2k_{\rm B} T}{A m_{\rm p}}} \frac{Z f_{\rm i}}{\Lambda}.
\label{eq5}
\end{equation}

\begin{figure}[t!]
\centering
\includegraphics[width=\linewidth]{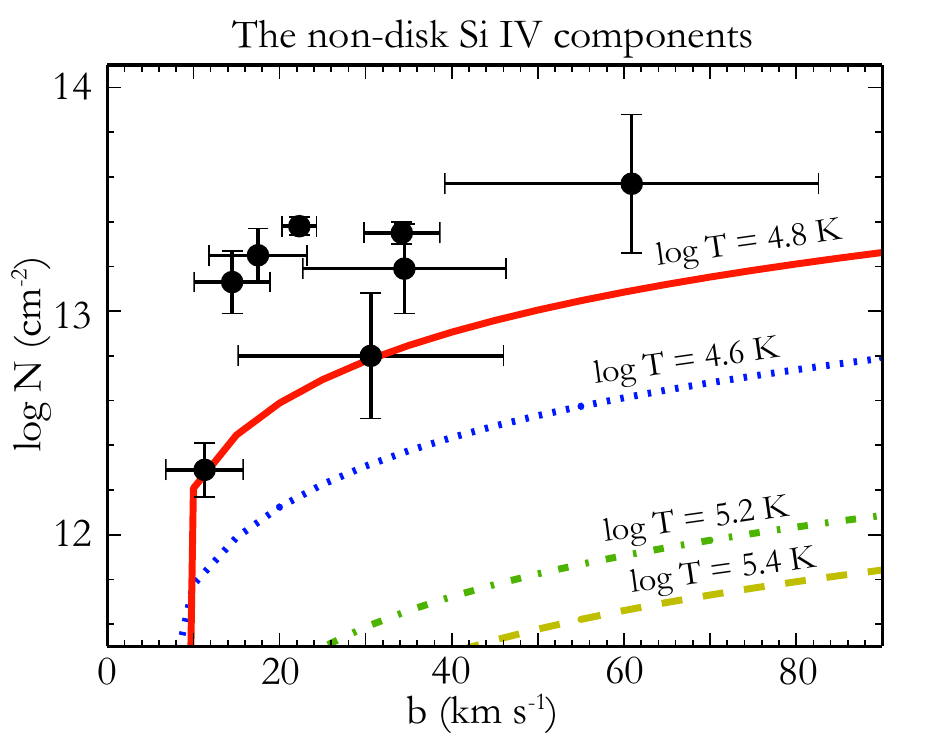}
\caption{The observed log $N$-$b$ correlation of the non-disk (disk-halo) components of \ion{Si}{4}. The colored curves show four CIE models derived using Eq \ref{eq5} assuming a metallicity of $Z$=0.1 $Z_{\odot}$ and at log $T$=4.6 K (blue dotted), 4.8 K (red solid), 5.2 K (green dash-dotted), and 5.4 K (yellow dash).}
\label{fig9}
\end{figure}

To test whether the non-disk \ion{Si}{4} can be produced under collisional ionization, we compare the observed log $N_{\rm SiIV}$ and $b$ with the theoretical relation from Eq \ref{eq5} in Fig \ref{fig9}. The $\Lambda$ and $f_i$ values are from \cite{Gnat07} assuming collisional ionization equilibrium (CIE) and Z=0.1 Z$_{\odot}$ is adopted which is within the metallicity range we derive in the following section. Fig \ref{fig9} shows that in both cases, log $N_{\rm SiIV}$ increases as $b$ becomes larger. For the CIE models, \ion{Si}{4} is mostly produced when log $T$ is 4.8 K, the temperature at which \ion{Si}{4} collisional ionization fraction reaches its maximum \citep{Gnat07}. However, even the highest production of \ion{Si}{4} from the CIE modeling is still $\sim0.5$ dex less than the vast majority of the observed data points.  

Our investigations indicate shocks and collisional ionization under CIE are unlikely the dominant mechanisms of producing \ion{Si}{4} at the disk-halo interface of M33. Photoionization is necessary to explain the large column density of \ion{Si}{4}. We proceed in the following section to discuss the role of photoionization in producing the observed \ion{Si}{4}. Note that non-equilibrium collisional ionization processes, such as turbulent mixing layers (\citealt{Slavin93, Kwak11}) and conductive interfaces (\citealt{Borkowski90, Gnat10}), could contribute in producing \ion{Si}{4} although current data is unable to distinguish among these scenarios.   

% \textbf{We note that there may exist unresolved \ion{Si}{4} components with smaller b values, which we missed during Voigt-profile fitting due to the limited {\it HST}/COS spectral resolution. \cite{Lehner11a} studied the \ion{Si}{4} ionization mechanism in MW's ISM with higher resolution spectra ($1.5-2.7\kms$) from the Space Telescope Imaging Spectrograph and found similar conclusions that \ion{Si}{4} with narrower profiles ($b\lesssim10\kms$) is generally photoionized or has radiatively cooled in non-equilibrium condition}.

\subsection{Photoionization Modeling and the Metallicity of the Disk-Halo Interface}
\label{sec6.2}

In this section, we consider the \ion{Si}{4} at the disk-halo interface being created by radiation from M33's starlight and extragalactic UV background (EUVB). We run two photoionization models for the M33 disk-halo interface using CLOUDY (v13.03; \citealt{Ferland98, Ferland13}), which is a radiative transfer code assuming the gas is in a plane-parallel slab with uniform density and in thermal equilibrium. We refer to the first model as ``pureEUVB", in which the gas is photoionized only by the assumed EUVB \citep{Haardt01}. It has an ionizing photon flux of $\Phi_{\gamma}=1.52\times10^5$ s$^{-1}$ cm$^{-2}$ \citep{Haardt01}, representing the minimum incident radiation field received by the disk-halo interface.
\footnote{But see \cite{Kollmeier14}, \cite{Shull15}, and \cite{Emerick15} for discussions of the discrepancy of photon production between the \cite{Haardt01} model and the \cite{Haardt12} model.}. 

The second model is called the ``EUVB+star" model, with ionizing flux from both EUVB and M33's starlight (SFR$=0.5\Msunyr$). We assume a photon escape fraction of 10\% and a gas slab distance of 5 kpc. The model has $\Phi_{\gamma}=7.86\times10^7$ s$^{-1}$ cm$^{-2}$, which is a combination of the EUVB \citep{Haardt01} and the stellar radiation using a L$_*$ galaxy's spectral energy distribution (SED) from Starburst99 \citep{Leitherer99} as in \cite{Werk14}. In this EUVB+star framework, the extent to which the galaxy's SED contributes to the photoionization rate depends on the escape fraction of ionizing photons $f_{\rm esc}$, the SFR of the galaxy, and the distance of the slab from the galaxy. The integrated flux from the EUVB+star model thus scales as ($f_{\rm esc}$ $\times$ SFR)/$d^{2}$. Given the uncertainties in estimating $f_{\rm esc}$ for M33 \citep{Hoopes00}, the SFR (\citealt{Engargiola03, Gratier10}), and the height of the accreting layer (Section \ref{sec5.1}), the EUVB+star model provides a rough but reasonable evaluation of the ionization condition of the disk-halo interface of M33. 

The models require an input of $N_{\rm H I}$ as a stopping condition. For the disk-halo gas, we obtain $N_{\rm HI}$ values by fitting Gaussian components to the AGES \ion{H}{1} 21-cm spectra within the integrated velocity ranges as indicated by the non-disk \ion{Si}{4} components. For each sightline, we show the non-disk \ion{Si}{4} and the corresponding \ion{H}{1} velocity ranges in blue in Fig \ref{fig3}. S6 shows the minimum contamination from the main disk emission while the others all have substantially large disk \ion{H}{1} wings entering the disk-halo velocity ranges. Given the importance of the \ion{H}{1} value, we only apply our pureEUVB and EUVB+star models to S6. The integrated non-disk \ion{H}{1} column density along S6 is $N_{\rm HI}=10^{19.0}$ cm$^{-2}$. This value may have uncertainties due to (1) the potential unresolved gas moving at similar velocities from behind star S6, and (2) \ion{H}{1} sub-structures within the beam of the radio observations (\citealt{Wakker01b, Tumlinson02, Welty12}). We repeat our CLOUDY modeling on a set of varying input log N$_{\rm HI}$ values from 17.0 to 21.0 cm$^{-2}$ (in step of 1.0 dex), and find that our derived [Si/H] (described below) values do not vary significantly for a range of log N$_{\rm HI}$ from 18.0 to 21.0 cm$^{-2}$. Therefore, the uncertainties related to $N_{\rm HI}$ do not affect our assessment of gas metallicity at the disk-halo interface of M33. 

The models predict \ion{Si}{4} column densities on $f_{\rm SiIV}$ - log $U$ - [X/H] grids, where $f_{\rm SiIV}$ is the gas fraction of \ion{Si}{4}, log $U$ the logarithm of the dimensionless ionization parameter $U$, and [X/H] the logarithm of the metallicity Z. [X/H] is also called the gas-phase abundance, which is normalized to a solar value: [X/H]$\equiv$log (N$_{\rm X}$/N$_{\rm H}$)-log (X/H)$_{\odot}$. For Silicon, we adopt (Si/H)$_{\odot}=10^{-4.49\pm0.04}$ from \cite{Asplund05}, which is the abundance reference of CLOUDY. The log $U$ and $f_{\rm SiIV}$ parameters of the disk-halo interface are solved by matching the predicted log $N_{\rm SiIV}$ with the observed one of S6, log $N_{\rm SiIV}$\footnote{S6 has two disk-halo \ion{Si}{4} components with one at 70.6$\kms$ and the other at 106.2$\kms$ as shown in Fig \ref{fig3} and Table \ref{tb2}. Here we use the sum of their column densities and calculate the error through error propagation. Note that the combined column density is within the maximum value as found from other individual components.}(=13.50$\pm$0.09 cm$^{-2}$). We apply the gas fraction $f_{\rm SiIV}$, $f_{\rm HI}$ to the observed log $N_{\rm SiIV}$ and log $N_{\rm HI}$ values to make ionization correction and calculate the metallicity [Si/H].

%\input t4.tex

%%%%%%%%%%%%%%%%%%%%

\begin{table}
\tabletypesize{\footnotesize}
\tablewidth{0pt}
\caption{CLOUDY Modeling using S6}
\begin{center}
\begin{tabular}{>{\footnotesize}c>{\footnotesize}c>{\footnotesize}c}
\hline
\hline
 & EUVB+star & pureEUVB \\
\hline
$\Phi_{\gamma}$\tablenotemark{a} (s$^{-1}$ cm$^{-2}$) & 7.86$\times$10$^7$  & 1.52$\times$10$^5$   \\
log $U$\tablenotemark{b}         & $[-2.4, -2.8]$     & $[-2.7, -3.3]$      \\
$n_{\rm H}$\tablenotemark{c} (cm$^{-3}$)  & $[0.7, 1.7]$ & $[0.003, 0.010]$  \\
$f_{\rm HI}$\tablenotemark{d}	 & $[0.026, 0.081]$   & $[0.042, 0.158]$   \\
$f_{\rm SiIV}$\tablenotemark{d}  & $[0.030, 0.007]$    & $[0.085, 0.018]$   \\
log (Si/H)\tablenotemark{e}      & $[-5.57, -4.45]$ & $[-5.82, -4.57]$    \\
${\rm [Si/H]}$\tablenotemark{f}  & $[-1.08, 0.04]$  & $[-1.33, -0.08]$   \\
\hline
\hline
\end{tabular}
\end{center}
\tablenotetext{a}{Ionizing photon flux. The pureEUVB value is from \cite{Haardt01}, and the EUVB+star one is a combination of the EUVB \citep{Haardt01} and the stellar ionization background \citep{Leitherer99}. }
\tablenotetext{b}{Logarithm of ionization parameter $U\equiv\Phi_{\gamma}/(cn_{\rm H})$ where $c$ is the speed of light.}
\tablenotetext{c}{Hydrogen number density $n_{\rm H}=\Phi_{\gamma}/(cU)$.}
\tablenotetext{d}{Gas fraction of \ion{H}{1} and \ion{Si}{4}, respectively. }
\tablenotetext{e}{Gas-phase abundance of Si with ionization correction.} 
\tablenotetext{f}{Gas-phase abundance of Si normalized to the solar value.}
\label{tb4}
\end{table}

%%%%%%%%Table 4%%%%%%%%%%%%%

Table \ref{tb4} shows the CLOUDY modeling results. The EUVB+star model predicts a higher log $U$ than the pureEUVB by $\sim$0.4 dex. The hydrogen volume density $n_{\rm H}$ of the EUVB+star model is 0.7--1.7 cm$^{-3}$, which is typical of \ion{H}{2} gas in the disk, while $n_{\rm H}$=0.003-0.01 cm$^{-3}$ from the pureEUVB model represents typical values of diffuse coronal gas \citep{Draine11}. Both models show small values of $f_{\rm HI}$, indicating that the hydrogen at the disk-halo interface is mostly ionized. The EUVB+star model shows [Si/H]=[-1.08, 0.04] and the pureEUVB model finds a similar range. Although the EUVB+star model has a much higher photon flux than the pureEUVB field, their SED slopes between the \ion{H}{1} ionization potential (1 Ryd) and the \ion{Si}{4}'s (2.5 Ryd) are similar (see Fig 13 in \citealt{Werk14}). This causes similar values of ($f_{\rm HI}/f_{\rm SiIV}$) and thus similar [Si/H] values despite the different $f_{\rm HI}$ and $f_{\rm SiIV}$ given by each model at a particular log U. 

The [Si/H] value we derive for the disk-halo interface of M33 is similar to that found by \cite{Lebouteiller06} for the \ion{H}{2} region NGC 604 (i.e., S5) using \ion{Si}{2} $\lambda$1020.7. Their value is [Si/X]=[-1.2, $\gtrsim$-0.2]. Since they use \ion{Si}{2} $\lambda$1020.7 with line centering at $\vlsr$=-245.4$\kms$ \citep{Lebouteiller06}, consistent with the systemic velocity of the gas disk at the position of S5 ($\vlsr$=-240.3$\kms$; Table \ref{tb1}), their measurement represents a [Si/H] value for M33's ISM. This indicates the disk-halo interface of M33 may have a metallicity similar to that of its ISM. In the next section, we calculate the metallicity of M33's ISM using the \ion{S}{2} abundance ratio. It should be noted that studies on MW's ISM show that Si is normally depleted by at least a factor of five in the ISM, and the depletion level may change as varying gas phases \citep{Savage96}. Overall, it is clear that M33's disk-halo interface has been previously enriched since its \ion{Si}{4} column density ($\langle N_{\rm SiIV}\rangle$=10$^{13.24}$ cm$^{-2}$) is comparable to both the disk's and the disk-halo interface gas of the MW that is known to be enriched ($\langle N_{\rm SiIV}\rangle$=10$^{13.17}$ cm$^{-2}$; \citealt{Shull09}). 

\subsection{The Metallicity of M33's ISM}
\label{sec6.3}

To assess the relative enrichment level of the disk-halo interface gas, we can compare the disk-halo metallicity with that of the ISM. Since the current M33's ISM metallicity measurements are mostly based on emission-line observations (e.g., \citealt{Crockett06, Magrini07a, Rubin08}), here we decide to derive the ISM abundance from our absorption-line data to avoid any systematic difference. We do not perform CLOUDY modeling for the ISM which would involve additional assumptions on SFR and incident radiation fields; instead, we use the abundance ratio [\ion{S}{2}/H] as S is not readily depleted onto interstellar dust \citep{Savage96} and the multiple transitions we detect are not saturated in our data. 

% \textbf{It is commonly assumed that the single-ionized \ion{S}{2} is the dominant form of the element S in the ISM \citep{Savage96}, while we note that a recent study of S ionization fraction indicates \ion{S}{3} could be as equally abundant as \ion{S}{2} in the Galactic warm ionized medium \citep{Howk12}.}

The total hydrogen column $N_{\rm H}$ in front of each star is obtained from the color excess E(B-V) as we have calculated in Section \ref{sec2.5}. Here we use the LMC-like log $N_{\rm H}$ values since M33 and LMC are more similar in terms of galaxy mass and metal content than the MW or SMC \citep{DOnghia15, Crockett06}. The reader should interpret the provided values with caution given the current limited knowledge of the actual  metal  content  of  M33's  ISM. If M33 is MW- (SMC-) like, the provided abundance ratios in Table \ref{tb5} would be higher (lower) by $\sim0.5$ (0.2) dex, while the relative abundance ratios [\ion{Fe}{2}/\ion{S}{2}] and [\ion{P}{2}/\ion{S}{2}] remain the same. Since the $N_{\rm H}$ value includes all the neutral hydrogen in front of a given star, we use the total ion column density of those lines that were fit with multiple kinematic components. This may slightly increase the derived ISM metallicity if some components are non-disk; however, this effect should be minor as ISM components usually dominate the absorption lines.

The [\ion{S}{2}/H] and [\ion{P}{2}/H] values in Table \ref{tb5} show that the metallicity of M33's ISM is $\sim0.1$ solar if M33 is LMC like. This is consistent with the derived metallicity range that we have found at the disk-halo interface with [Si/H]=[-1.08, 0.04], indicating that the disk-halo interface of M33 is as similarly metal-enriched as the ISM. [\ion{Fe}{2}/H] shows a mean value $\sim1.0$ dex lower than the other two, suggesting Fe is strongly depleted in M33's ISM. We additionally calculate the [\ion{P}{2}/\ion{S}{2}] and [\ion{Fe}{2}/\ion{S}{2}] abundance ratios. These values do not require the input of $N_{\rm H}$, thus provide unbiased measurements of the abundance pattern of M33's ISM. Fe again shows a relative depletion with respect to S.

%\input t5.tex

%%%%%%%%%%%%%%%%%
\begin{table}
\footnotesize
\caption{Metallicity of M33's ISM from Absorption Line}
\begin{center}
%\begin{tabular}{>{\footnotesize}c>{\footnotesize}c>{\footnotesize}c>{\footnotesize}c>{\footnotesize}c>{\footnotesize}c}
\begin{tabular}{X{0.2cm}X{1.0cm}X{1.0cm}X{1.2cm}X{1.8cm}X{1.3cm}}
\hline
\hline
ID & [\ion{S}{2}/H]\tablenotemark{a} & [\ion{P}{2}/H]\tablenotemark{a} & [\ion{Fe}{2}/H]\tablenotemark{a} & [\ion{Fe}{2}/\ion{S}{2}] \tablenotemark{b}& [\ion{P}{2}/\ion{S}{2}]\tablenotemark{b}\\
\hline
\hline
S1 & -1.0 & -0.8  & -1.9 & -0.9$\pm$0.2 & 0.1$\pm$0.4\\
S2 & -1.3 & -1.1  & -2.2 & -0.9$\pm$0.2 & 0.1$\pm$0.2\\
S3 & -1.0 & -0.9  & -2.0 & -1.0$\pm$0.1 & 0.1$\pm$0.1\\
S4 & -1.4 & -1.1  & -2.6 & -1.2$\pm$0.1 & 0.3$\pm$0.1\\
S5 & -1.2 & -1.2  & -2.2 & -1.0$\pm$0.1 & 0.0$\pm$0.1\\
S6 & -1.2 & $<$-1.4  & -2.1 & -0.9$\pm$0.2 & $<$-0.1\\
S7 & -1.5 & -1.3  & -2.5 & -1.0$\pm$0.1 & 0.2$\pm$0.1\\
\hline
\hline
\end{tabular}
\end{center}
\tablenotetext{a}{Gas-phase abundance.  [X/H]$\equiv$log($N_{\rm X}$/$N_{\rm H}$)-(X/H)$_{\odot}$ where X represents species \ion{S}{2}, \ion{P}{2} and \ion{Fe}{2}, respectively. For the solar values, we adopt (S/H)$_{\odot}$=10$^{-4.86\pm0.05}$, (P/H)$_{\odot}$=10$^{-6.64\pm0.04}$, and (Fe/H)$_{\odot}$=10$^{-4.55\pm0.05}$ \citep{Asplund05}. $N_{\rm H}$ is the LMC-like values in Table \ref{tb3}. If M33 is MW (SMC) like, the provided abundance ratios would be higher (lower) by $\sim0.5$ (0.2) dex, while the relative abundance ratios [\ion{Fe}{2}/\ion{S}{2}] and [\ion{P}{2}/\ion{S}{2}] remain the same. The reader should interpret the provided metallicities with caution given the current limited knowledge of the actual metal content of M33's ISM. }
\tablenotetext{b}{The relative abundance ratio, defined as [X/\ion{S}{2}]$\equiv$log($N_{\rm X}$/$N_{\rm SII}$)-(X/S)$_{\odot}$ where X represents \ion{Fe}{2} and \ion{P}{2}, respectively. }
\label{tb5}
\end{table}

%%%%% table 5 %%%

We note that our S and P metallicity is consistently lower than those values based on emission-line measurements. For example, \cite{Rubin08} found [S/H] $\sim-0.5$ dex and \cite{Crockett06} indicated [O/H] $\sim-0.4$ dex, both based on emission-line observations of \ion{H}{2} regions in M33. This may reflect a systematic difference between emission-line- and absorption-line-based metallicity measurements. We do not attempt to analyze M33's ISM further as the focus of this work is the disk-halo gas. However, we tabulate the data in Table \ref{tb5} for researchers who are interested in the metal content of M33's ISM.

%---------------------------- Discussion -----------------------
\section{Discussion}
\label{sec7}

Using UV-bright M33 stars as background targets, we detect an unambiguous accreting ionized gas flow towards the disk of M33. The ionized gas flow is multi-phase, as it is consistently seen in the warm ion \ion{Si}{4} and it is also found in the cool ions (\ion{Fe}{2}, \ion{S}{2}, and \ion{P}{2}) along some sightlines. It is also possibly present in the saturated \ion{Si}{2}, \ion{Si}{3}, and \ion{C}{2} lines. Section \ref{sec5} shows that the inflow can be well explained by an accreting layer with an inflow velocity of 110$_{-20}^{+15}\kms$ and at a z-height of 1.5$_{-1.0}^{+1.0}$ kpc from the galactic plane. In Section \ref{sec6}, we find that the disk-halo interface gas of M33 is metal-enriched. An estimate of the metallicity of this accreting layer is [Si/H]=[-1.08, 0.04], consistent with the metallicity of M33's ISM measured via similar absorption-line methods.  In this section, we calculate the mass and the accretion rate of the accreting layer and discuss its origin(s). We discuss the potential existence of outflows from M33 in the final part of this section.

\subsection{Accretion Mass and Gas Accretion Rate}
\label{sec7.1}

The kinematic modeling in Section \ref{sec5} predicts an accreting layer with an inflow velocity of $v_{\rm acc}=110\kms$ at height of $z_{\rm acc}=1.5$ kpc. Assuming a cylindrical geometry for the accreting layer with height $z_{\rm acc}$ and radius $R_{\rm acc}$, the total mass of the detected accreting layer is, 
\begin{equation}
\begin{split}
M_{\rm acc} & =1.3\times(\pi R_{\rm acc}^2 z_{\rm acc}) m_{\rm H} N_{\rm H}/z_{\rm acc} \\
            & =1.3\pi R_{\rm acc}^2 m_{\rm H} N_{\rm SiIV} [\frac{\rm Si}{\rm H}]_{\odot}^{-1} Z^{-1} f_{\rm SiIV}^{-1} \\
            & \approx 3.9\times10^7 \Msun \\ 
            & (\frac{R_{\rm acc}}{7\ {\rm kpc}})^2 (\frac{N_{\rm SiIV}}{10^{13.2} {\rm cm^{-2}}})(\frac{Z}{0.2\ Z_{\odot}})^{-1}(\frac{f_{\rm SiIV}}{0.1})^{-1}
\end{split}
\label{eq:macc}
\end{equation}
where 1.3 accounts for the helium mass using a primordial He/H ratio of 1/12 and $m_{\rm H}$ is the proton mass. For $R_{\rm acc}$, we adopt the maximum inclination-corrected galactocentric distance $R_{\rm G}\sim7$ kpc (Table \ref{tb1}) assuming the accreting layer covers the whole star-forming disk. The term $\pi R_{\rm acc}^2 z_{\rm acc}$ is the total volume of the layer in front of the stars, and $m_{\rm H} N_{\rm H}/z_{\rm acc}$ is the mass density. Since large uncertainties are associated with the estimation of $N_{\rm H}$ for the disk-halo gas, here we substitute $N_{\rm H}$ with the measured $N_{\rm SiIV}$: $N_{\rm H}=N_{\rm SiIV} (\frac{\rm Si}{\rm H})_{\odot}^{-1} Z^{-1} f_{\rm SiIV}^{-1}$. We adopt $N_{\rm SiIV}=10^{13.2}$ cm$^{-2}$ which is the mean \ion{Si}{4} column density of the disk-halo gas, $Z\sim0.2$ $Z_{\odot}$ which is within the metallicity derived in Section \ref{sec6.2}, and $f_{\rm SiIV}\sim0.1$ which is the maximum gas fraction of \ion{Si}{4} from both ionization fields (Table \ref{tb4}). Note that our calculation is conservative and the actual accretion mass and the accretion rate could be higher if we use a smaller gas fraction $f_{\rm SiIV}$ value. If a symmetric accreting layer exists on the other side of the disk, the total mass of accreting material will double from the value above to $\sim7.8\times10^7\Msun$. 

The closest observed analogs to this accreting layer are the denser \ion{H}{1} IVCs ($d\leq3-4$ kpc; \citealt{Albert04}) and the H$\alpha$ Reynolds layer ($1-2$ kpc; \citealt{Reynolds93, Haffner03}) in the MW, or the \ion{H}{1} extraplanar gas (e.g., \citealt{Fraternali02, Oosterloo07, Gentile13}) and diffuse ionized gas (e.g., \citealt{Collins00, Heald06}) in external galaxies. The M33 accreting layer is much less massive than these (at least a factor of ten less than the \ion{H}{1} extraplanar gas in other galaxies; e.g., NGC 2403; \citealt{Fraternali02}) and is clearly infalling, while these observed analogs are only known to be lagging in rotation. It is interesting that our Accreting Layer model (Section \ref{sec5.1}) is not sensitive to halo lagging since the lagging rotational velocity is found along the vertical direction out to a few kpc in both \ion{H}{1} and H$\alpha$ in the MW and external galaxies (e.g., \citealt{Heald06b, Levine08}). Future observations that are capable of observing more direct analogs to our accreting layer would be useful, in particular, observations that can kinematically resolve the extraplanar gas and identify inflow motions. 

We derive the accretion rate for the accreting layer on the front side of the disk,
\begin{equation}
\begin{split}
\dot{M} & = M_{\rm acc} v_{\rm acc}/z_{\rm acc}  \\
        & \approx 2.9\Msunyr (\frac{M_{\rm acc}}{3.9\times10^7 \Msun}) (\frac{v_{\rm acc}}{110\ \kms})(\frac{z_{\rm acc}}{1.5\ {\rm kpc}})^{-1}.
\end{split}
\label{eq:chi2}
\end{equation}
Again, if there is a symmetric layer on the other side of the disk, the total accretion rate will be $\sim5.8\Msunyr$. This accretion rate is much higher than the M33's SFR ($\lesssim0.5\Msunyr$;  \citealt{Engargiola03, Gratier10}). In the next section, we discuss two potential scenarios that may account for the large accretion rate. We note that the covering fraction of the gas may play a critical role in evaluating the accreting mass and the accretion rate. In our calculation, we assume a 100\% coverage since all seven sightlines have detected \ion{Si}{4} inflow. However, since our sightlines are all targeted at star-forming regions, we do not have the comparison with non-star-forming regions in M33 disk. The $M_{\rm acc}$ and $\dot{M}$ would be lower or higher if the accreting inflow has lower or higher column density in non-star-forming regions. Future observations using QSO sightlines through non-star-forming regions may help to evaluate the structure of the accreting gas. 

\subsection{The Origin of the Ionized Gas Inflow}
\label{sec7.2}

As discussed in Section \ref{sec6}, the disk-halo gas of M33 is metal-enriched. This enrichment could result from some disk-wide gas mixing processes as we find a 100\% detection rate of infalling \ion{Si}{4} and our sightlines have sampled the disk rather evenly, including both the very center of the disk and its star-forming outskirts. We propose two scenarios that could account for the detected gas inflows at the disk-halo interface of M33.

First, the accreting layer may be the cooling stage of the galactic fountain model \citep{Shapiro76}. The model starts with hot gas in the ISM losing pressure support from its surrounding cool medium and streaming upward and radially outward to a height of a few kpc (\citealt{Bregman80, Houck90}). Then it condenses into cold clouds via non-linear instability \citep{Joung12a}. Once the cold clouds can no longer feel pressure support from the hot corona, they fall downward and inward due to gravity and centrifugal forces and can dissolve into the warm ionized medium after travelling several kpc (\citealt{Heitsch09, Joung12a}). 

\cite{Bregman80} ran a 2D hydrodynamic simulation incorporating the galactic fountain scenario to explain the existence of HVCs in the MW halo. He found that if the cold clouds fall ballistically, their vertical highest velocities could reach -80$\kms$ and up to -150$\kms$. Similar results are found if we use the terminal velocity equation from \cite{Benjamin97} to calculate the terminal velocities of free-falling clouds at the disk-halo interface. This is consistent with the high velocity $v_{\rm acc}=110\kms$ found in our Accreting Layer model. 

A number of authors (e.g., \citealt{Fraternali06, Fraternali08, Marinacci11, Marasco12}) suggest that the cold fountain clouds would seed the condensation of the hot corona and trigger a net inflow of halo gas. The total inflow includes the fountain return, and the ambient halo material that contributes $10-20$\% of the inflow mass and feeds the star formation. Taking into account the fountain return and the newly accreted halo gas, the total accretion rate should be $\sim5-10$ times the SFR of the galaxy. If we apply this fountain model to M33, which has a SFR $\lesssim0.5\Msunyr$, one would expect a total accretion rate of $\dot{M}\lesssim2.5-5.0\Msunyr$, consistent with the number we derive in Section \ref{sec7.1}. The galactic fountain model predicts changes of gas phases at different circulation stages; this is likely our case as warm and cool ions are both found with accreting velocities. However, we are unable to fully address whether the accreting flow consistently bears a cool-phase gas as the non-disk components in \ion{Fe}{2}, \ion{S}{2}, and \ion{P}{2} are only detected along a few sightlines and the \ion{Si}{2}, \ion{Si}{3}, and \ion{C}{2} lines are saturated.

Alternatively, the accreting gas could be associated with stripped material raining back down that was pulled from the disk of M33 during a close passage with M31. The M31-M33 interaction is hinted by its large warps, diffuse \ion{H}{1} features, and extended stellar disk (\citealt{vanderMarel08, Putman09, McConnachie09}). \cite{Heitsch09} showed that if falling from 10 kpc, the maximum velocity a free-falling cloud could reach is $\lesssim150\kms$, which is consistent with the accretion velocity $v_{\rm acc}=110\kms$ of our fast accreting layer. In this case, the large accretion rate of $\sim2.9\Msunyr$ would be temporary and result in an increase in star formation in the disk, but may not have a long-term influence. 

Apart from these two scenarios, there may exist other kinematic processes that have not been accounted for in order to fully understand M33's accretion. For future work, more sophisticated kinematic models incorporating radial gas flows (e.g., \citealt{Schmidt16}) would be helpful to disentangle the degeneracy of different gas motions. Detailed galactic fountain modeling that takes into account the distribution of local star-forming regions in the disk would shed light upon the influence of stellar activities in shaping the disk-halo interface of M33. Observationally, more sightlines using bright stars in the disk and background QSOs through non-star-forming regions will help to address whether there exists a correlation between gas inflows and local star formation.

\subsection{Outflows from M33}
\label{sec7.3}

Galactic outflows driven by stellar and/or AGN feedback are ubiquitously observed in star-forming and starburst galaxies both in the local universe and at higher redshifts (e.g., \citealt{Lehner09, Heckman00, Shapley03, Chen10, Weiner09, Martin12, Rubin14}). Winds have been observed in multiple phases traced by CO emission, H$\alpha$ emission, UV and optical absorption lines (e.g., \ion{Mg}{2}, \ion{Fe}{2}, \ion{Na}{1}) and X-ray (e.g., \citealt{Martin99, Walter02, Strickland04a, Strickland04b}). Galactic outflows are generally found moving at several hundred$\kms$ and can be seen out to $\sim1000\kms$ (e.g., \citealt{Weiner09, Coil11, Rubin14}). 

One may expect the presence of such galactic outflows in M33 since it is actively forming stars (e.g., \citealt{Engargiola03, Gratier10}) and has a relatively high SFR surface density (\citealt{Heyer04, Magrini07b}) that could potentially drive winds out of the disk (e.g., \citealt{Rubin14}). In addition, the metallicity of M33's ISM is abnormally low in terms of its stellar mass (3--6$\times$10$^9\Msun$; \citealt{Corbelli03}) according to the mass-metallicity relation \citep{Tremonti04}, also suggesting the presence of outflows in M33 to efficiently remove metals from its ISM. Accordingly, we now discuss whether the M33 COS data show any evidence for ionized gas outflows along our seven sightlines. 

\subsubsection{Tentative Detection of Slow Winds}
\label{sec7.3.1}

We detect weak blueshifted absorption components ($-40\lesssim v_{\rm d}\lesssim 0\kms$) in low and/or warm ions along our sightlines (see Fig \ref{fig3}). As shown in Table \ref{tb2}, the mean column densities for these blueshifted components are $N_{\rm S\ II}=10^{15.17}$ cm$^{-2}$, $N_{\rm P\ II}=10^{13.52}$ cm$^{-2}$, $N_{\rm Fe\ II}=10^{14.26}$ cm$^{-2}$, and $N_{\rm Si\ IV}=10^{13.43}$ cm$^{-2}$, and their mean Doppler widths are $b\sim25-40\kms$. We find that these blueshifted components could be due to the expansion of local \ion{H}{2} regions into M33's ISM, since both their $b$ values and measured velocities are consistent with the large velocity dispersions found in the ionized gas of \ion{H}{2} regions ($\sigma\sim20-30\kms$; \citealt{Kam15}) and in the \ion{H}{1} 21-cm emission across the M33 disk ($\sigma\sim18.5\kms$; \citealt{Putman09}). In particular, the giant \ion{H}{2} region NGC 604 (i.e., S5) has been found to be undergoing multiple blowouts \citep{Tenorio00}, consistent with the blue-shifted components we detect for S5. We note that these blueshifted components have velocities that are close to our CalCOS spectral uncertainty ($14-19\kms$), thus they only tentatively exhibit evidence for slow winds. In general, these slow-velocity features indicate gas activities that are closely related to M33's ISM.   

\subsubsection{Potential Fast Outflows at $v_{\rm d}\lesssim-100\kms$?}
\label{sec7.3.2}

As indicated in Section \ref{sec3}, we find strong absorption signatures in \ion{C}{2}, \ion{Si}{2}, and \ion{Si}{3} along all our available sightlines at negative high velocities of $-200\lesssim v_{d}\lesssim-100\kms$ (see Fig \ref{figa1}); however, such absorption is not found in \ion{Si}{4}, \ion{P}{2}, \ion{S}{2}, or \ion{Fe}{2}. In this section, we discuss the possible origins for these negative high-velocity \ion{C}{2}, \ion{Si}{2}, and \ion{Si}{3} absorbers, while in the following section (\ref{sec7.3.3}), we address the non-detection of relevant absorption in \ion{Si}{4}, \ion{P}{2}, \ion{S}{2}, or \ion{Fe}{2}. 

The \ion{C}{2}, \ion{Si}{2}, and \ion{Si}{3} absorbers could potentially be fast outflows that have originated from the disk of M33, as their velocities are broadly consistent with those measured in extragalactic down-the-barrel studies (e.g., \citealt{Rubin14}). If such outflows indeed exist in M33, they can only be driven by stellar winds or supernova explosions (such as those in LMC; \citealt{Lehner09}) as M33 does not have a central massive black hole \citep{Merritt01}. This would be an unusual detection however, as these absorbers generally have narrow line widths (FWHM$<100\kms$) compared to the ionized galactic outflows with FWHM values of $\sim200-500\kms$ found in star-forming regions (e.g., \citealt{Chisholm16}).

On the other hand, the \ion{C}{2}, \ion{Si}{2}, and \ion{Si}{3} absorbers may have other origins. First, the absorbers may be part of the northern warp of M33 that folds back over the disk thus to intercept our sightlines, given that M33 has a large extended northern \ion{H}{1} warp with velocity up to $\vlsr\sim-300\kms$ (\citealt{Rogstad76, Corbelli97, Putman09}). Second, this low-ionization state absorption may be due to stripped debris from M33 if the galaxy orbited close to M31 in the past several Gyrs (\citealt{Putman09, McConnachie09}). Third, even though their velocities differ in the M33's $v_{\rm d}$ frame, these \ion{C}{2}, \ion{Si}{2}, and \ion{Si}{3} absorbers have nearly constant $\vlsr$($\sim-350\kms$), suggesting that they may come from the ionized extension of a nearby high-velocity cloud -- the Wright's Cloud (\citealt{Wright79, Braun04, Putman09}). In a deep \ion{H}{1} 21-cm emission map of the cloud \citep{Keenan15}, the ionized gas absorption appears well-aligned in position space and, its velocity matches the observed gradient of the cloud. 

%However, we do not detect similar absorption in cool ions \ion{Fe}{2}, \ion{S}{2}, \ion{P}{2} or the warm ion \ion{Si}{4}, which we discuss in Section \ref{sec7.3.3}.

In all, we find that the \ion{C}{2}, \ion{Si}{2}, and \ion{Si}{3} absorbers at $-200\lesssim v_{d}\lesssim-100\kms$ ($\vlsr\sim-350\kms$) could be a mixture of multiple sources given the complex surrounding environments. Their presence certainly helps to provide a more complete picture of ionized gas environment near M33, thus an intricate analysis on its origin should be carried out, which is beyond the scope of this work. We will extend our investigation on the origin(s) of these negative high-velocity \ion{C}{2}, \ion{Si}{2}, and \ion{Si}{3} absorbers in a separate paper (Zheng et al., 2017, in prep.).

\subsubsection{Covering Fraction of the Potential Fast Outflows}
\label{sec7.3.3}

If the blue-shifted absorption seen in \ion{C}{2}, \ion{Si}{2}, and \ion{Si}{3} have origins other than outflowing gas, then M33 exhibits a lack of fast outflows in low-ionization states. This scenario is also supported by the non-detections of fast outflows in \ion{Fe}{2}, a frequently detected ion in down-the-barrel wind studies (e.g., \citealt{Rubin14}), as well as in \ion{P}{2}, \ion{S}{2}, or \ion{Si}{4}. In this section, we discuss the possible reasons for the non-detections in these ions, and use the non-detections to constraint the upper limit of the covering fraction of the potential outflows.

First, we find that the non detection of outflows among these ions cannot be explained by the potential geometry of the outflows or the inclination of M33's disk. Generally speaking, outflows having collimated and/or bi-conical geometry are easier to observe in face-on galaxies (e.g., \citealt{Martin12, Rubin14}); thus, the inclination of M33 of $56^{\degree}$, could preclude the detection of outflows in our study. However, we observed seven sightlines that evenly sampled different areas of the disk. Three of them are within two kpc from the galactic center. If such collimated outflows have a large covering fraction (as discussed below), we would have been able to detect them even though M33 is moderately inclined. 

Second, the non-detection is unlikely due to data's sensitivity limit. Our {\it HST}/COS spectra are sensitive to \ion{Fe}{2} column density above 10$^{13.5}$ cm$^{-2}$ while the typical $N_{\rm FeII}$ value found for the galactic outflows in star-forming galaxies at redshift $\sim0.5$ is $\gtrsim10^{14.0}$ cm$^{-2}$ \citep{Rubin14}. The fast outflows in M33 could potentially be in a hotter phase than we can detect as the disk could blow out hot and diffuse gas as pictured in the galactic fountain model (see Section \ref{sec7.2}). However, this speculation is inconsistent with most observations that commonly find outflows in colder phases traced by cool ions such as \ion{Fe}{2} (e.g., \citealt{Chen10, Coil11, Rubin14}).

Third, the small aperture size of our COS observations ($\sim$10 pc) would not cause the non-detection if the outflows have covering fraction (C$_f$) of 1.0 over the disk. However, it is possible that the outflows from M33 have a substantially low covering fraction C$_f$ thus elude our detections. With the seven non-detections in our COS data, we can assess the upper limit to C$_f$. In this framework, the likelihood of a non-detection along a given sightline is $1-{\rm C}_f$. Assuming that the seven non-detection events are independent, which is reasonable given the typical projected separation between sightlines of $\sim$2 kpc, the probability of seven non-detections is then ($1-{\rm C}_f$)$^7$. To constrain the upper limit to the covering fraction at 95\% confidence level, we then expect ($1-{\rm C}_f$)$^7\geq(1-0.95)$. Thus, we find that the covering fraction of outflows from M33 with $N_{\rm FeII}>10^{13.5}$ cm$^{-2}$ is constrained to be C$_f\leq0.35$ at a 95\% confidence. This suggests the covering fraction of outflows (in \ion{S}{2}, \ion{S}{2}, \ion{P}{2}, and \ion{Si}{4}) from M33 have to be less than 35\% if they indeed exist. 

The derived covering fraction for M33's outflows lies at the low-end C$_f$ ranges (0.24--0.98) of the ubiquitous outflows found in star-forming galaxies at z $\sim$ 0.5 \citep{Rubin14}. In their sample, the galaxies with winds exhibit a mean covering fraction of C$_f$=0.59$\pm$0.17. The mean C$_f$ would be lower (0.37$\pm$0.32) if we include those galaxies that do not have wind detection and assign them C$_f$ values of 0. Similar results are found if we only consider galaxies with inclination of 50-60$^{\degree}$ (M33 has $i$ = 56$^{\degree}$). Therefore, the low covering fraction of the potential outflows from M33 is consistent with observations of outflows from star-forming galaxies based on larger data sample. 

Note that this direct comparison may be complicated by the varying column density limits of \cite{Rubin14} at different covering fraction. Beside the above statistical argument, it is worth noticing that our sightlines are through major star forming regions in the disk where one would expect to detect galactic outflows if they were present. We suggest that for future work an increased number of sightlines through the disk at different locations or an increase of aperture size will help to increase the outflow detection probability or solidify that M33 does not host galactic outflows.

%---------------------------- Conclusion -----------------------
\section{Conclusion}
\label{sec8}

We present a UV absorption-line study of the ionized gas inflow at the disk-halo interface of M33. In contrast to common approaches of using background QSOs, we choose a sample of seven UV-bright stars in the disk of M33. This allows us to unambiguously identify gas inflow onto the disk. Our main results are summarized as follows.

We detect consistently redshifted non-disk \ion{Si}{4} absorption along all of our sightlines that extend evenly across the disk of M33 (Section \ref{sec3}). Their centroid velocities $v_{\rm d}$ range from +39$\kms$ to +106$\kms$ with a mean value of +67$\kms$ (or +67/cos56$^{\degree}$ $\sim+120\kms$ if corrected for inclination), suggesting that there exists a disk-wide, warm-ionized gas accreting toward M33. This accreting flow is most likely multi-phase, since similar non-disk components can also be observed in cool ions (\ion{Fe}{2}, \ion{S}{2}, and \ion{P}{2}) along some sightlines. They may also commonly exist in saturated ion lines (\ion{C}{2}, \ion{Si}{2}, and \ion{Si}{3}). The mean column density of the non-disk ionized gas is $\langle N_{\rm SiIV}\rangle$=10$^{13.24\pm0.21}$ cm$^{-2}$. We find that shocks and collisional ionization under CIE cannot account for the total non-disk \ion{Si}{4} column and, we conclude that the majority of \ion{Si}{4} is produced via photoionization (Section \ref{sec6.1}).  

To interpret the consistently redshifted non-disk \ion{Si}{4} absorption components, we construct two kinematic models - the Accreting Layer model and the M33 Halo Cloud model (Section \ref{sec5}). We find the Accreting Layer model provides the best explanation of the observed non-disk \ion{Si}{4} velocities if the layer accretes toward the disk at $v_{\rm acc}=$110$_{-20}^{+15}\kms$ at a height of $ z_{\rm acc}=$1.5$_{-1.0}^{+1.0}$ kpc. The mass of the accreting layer on the near side of the disk is 3.9$\times$10$^7\Msun$, and the accretion rate is 2.9$\Msunyr$ (Section \ref{sec7.1}). If symmetric accreting flows exist on both sides of the disk, the total accretion mass is 7.8$\times$10$^7\Msun$ and the accretion rate is 5.8$\Msunyr$. 

We find an abundance ratio of [Si/H]=[-1.08, 0.04] for the accreting gas at the disk-halo interface of M33. Although there are uncertainties that make it somewhat unclear if M33's disk-halo gas has a higher or lower metallicity than its ISM ([\ion{S}{2}/H] $\sim$ -1.0 dex), its \ion{Si}{4} column density ($10^{13.24\pm0.21}$ cm$^{-2}$) is comparable to the M33 disk gas and the enriched disk-halo interface of the MW ($10^{13.17}$ cm$^{-2}$; \citealt{Shull09}).

We invoke two processes in the discussion of the origin of the metal-enriched, fast-accreting gas at the disk-halo interface of M33. First, we find that a galactic fountain model is able to explain the metal enrichment via gas recycling, the high inflow velocity by ballistic motions, and the large accretion rate by bringing back more material from the hot corona. Second, we speculate that the accreting layer could include free-falling material that was previously stripped from the disk of M33 as it orbited close to M31. In this case, the large accretion rate would be temporary. More sophisticated modeling is needed to better understand the gas accretion at the disk-halo interface of M33. 

We detect negative high-velocity absorption in \ion{C}{2}, \ion{Si}{2}, and \ion{Si}{3}. Their origin is so far unclear -- M33's fast outflows, M33's \ion{H}{1} warp or a nearby HVC, could all be possible. We defer the detailed analysis of the origins of these ion absorbers in a future paper (Zheng et al. 2017, in prep.). We do not have detection at negative high velocities in \ion{Fe}{2}, \ion{S}{2}, \ion{P}{2}, or \ion{Si}{4} along any of our sightlines. This is unlikely due to geometric effects with the outflows being collimated and bi-conical, as our sightlines extend across the star-forming disk. We suggest that the covering fraction of the potential galactic outflows has to be less than 35\% at a 95\% confidence level (at a $N_{\rm FeII}>10^{13.5}$ cm$^{-2}$ sensitivity limit) if M33 indeed hosts such fast-moving winds. 

Our study of M33 is one of the first to clearly reveal the existence of an ionized gas inflow onto the disk of a galaxy beyond the MW. Our analysis provides a better understanding of gas accretion in the vicinity of a galaxy disk; however, there are several ways we can further our understanding in the future. 1) The sample of sightlines should be enlarged to assess the clumpiness and sub-structure of the inflowing gas. 2) It would be ideal to have QSO sightlines that do not go through \ion{H}{2} regions and/or spiral arms to determine the effects of local star formation activities. 3) With only COS/G130M grating, a critical ion \ion{C}{4}, which traces collisional processes, is missed.  It would be helpful to use the G160M grating to expand the detected line library and form a clearer picture of the ionization state of the gas. 4) Finally, in the future more detailed models that incorporate different kinematic processes would be of great help in interpreting the observed gas motions. 

{\it Acknowledgement}. We thank the anonymous referee for the detailed comments to improve this work. We appreciate B.P. Wakker's generosity of providing us the co-added spectra of S7 ran with his program. We thanks C. Liang, H.W. Chen and A.J. Fox for discussions on {\it HST}/COS wavelength calibration and comments on the draft, J. Tumlinson, J.M. Shull and C.W. Danforth for discussions on \ion{Si}{4} and \ion{O}{6} line profiles, N. Lehner for questioning and reviewing Section \ref{sec4}, K.H.R. Rubin for discussions on our non-detection of galactic-scale outflows in \ion{Fe}{2}, and J.X. Prochaska, Filippo Fraternali and David J. Helfand for useful comments on the draft. J. Werk and Y. Zheng gratefully acknowledge the peaceful and beautiful setting offered by the Esalen institute during the drafting of this manuscript.

This work is based on observations made with the NASA/ESA Hubble Space Telescope (program ID: 13706). Support for HST-GO-13706 was provided by NASA through a grant from the Space Telescope Science Institute (STScI). STScI is operated by the Association of Universities for Research in Astronomy, Inc., under NASA contract NAS5-26555. Some of the data presented in this paper were obtained from the Mikulski Archive for Space Telescopes (MAST). Support for MAST for non-HST data is provided by the NASA Office of Space Science via grant NNX13AC07G and by other grants and contracts. We also acknowledge support from the National Science Foundation under Grant No. AST-1410800. J.E.G. Peek and J.K. Werk were supported in part by Hubble Fellowship grants 51295 and 51332, respectively, provided by NASA to STScI. 

This research made use of the IPython package \citep{Ipython07}, matplotlib \citep{Hunter07} which is a Python library for publication quality graphics, and Astropy \citep{Astropy13} which is a community-developed core Python package for Astronomy. CalCOS is a product of the Space Telescope Science Institute, which is operated by AURA for NASA. This research also made use of the SIMBAD database, operated at CDS, Strasbourg, France. The M33 image is made with the NASA Galaxy Evolution Explorer. GALEX is operated for NASA by the California Institute of Technology under NASA contract NAS5-98034.

%---------------------------- References -----------------------
\bibliographystyle{apj}
\bibliography{Zheng16_M33ACC}

\appendix

\section{Properties of Saturated Ions}
\label{appA}

\begin{figure*}[t]
\includegraphics[width=\textwidth]{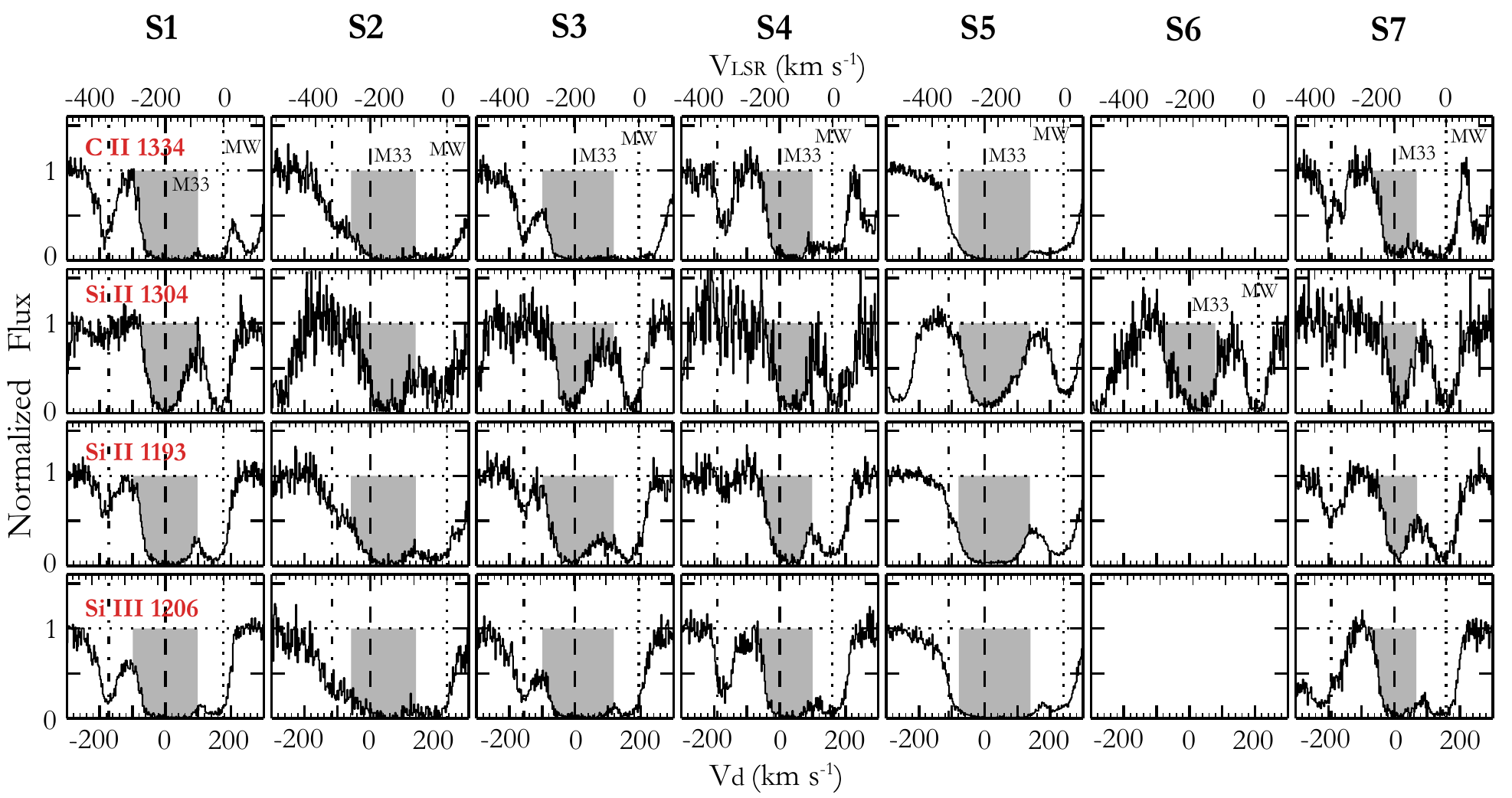}
\caption{The continuum-normalized {\it HST}/COS spectra of \ion{C}{2} $\lambda$1334, \ion{Si}{2} $\lambda\lambda$1304, 1193, and \ion{Si}{3} $\lambda$1206. Velocity on the bottom X axis is shown with respect to the gas disk of M33 at the position of each sightline. The velocity on the top X axis is shown in the LSR frame. The vertical dashed line show the $v_{\rm d}=0\kms$, which is the systemic velocity of the M33 disk at the position of the corresponding sightline. The dotted line around $v_{\rm d}> +150\kms$ ($\vlsr=0\kms$) shows the MW absorption, while the dash-dotted line at $v_{\rm d}\sim-200\kms$ ($\vlsr=-350\kms$) indicates the negative high-velocity absorption that is discussed in Section \ref{sec7.3}. The highlighted gray shades show the velocity ranges we use to calculate the M33-associated ion column density via the AOD method. Note that we do not attempt to normalize the \ion{C}{2}, \ion{Si}{2}, and \ion{Si}{3} spectra of S6 (except for \ion{Si}{2} $\lambda$1304) since they are strongly affected by stellar lines.}
\label{figa1}
\end{figure*}

%\input t6.tex
%%%%%%%%%%%%%%%%%%%%%
\begin{table}
\renewcommand{\arraystretch}{0.9}
\footnotesize
\tablewidth{\columnwidth}
\caption{AODM: Saturated Ions}
\begin{center}
\begin{tabular}{X{0.5cm}X{1.8cm}X{1.3cm}X{1.4cm}X{1.5cm}}
\hline
\hline
 & $v_{\rm min}$, $v_{\rm max}$ & log N$_{\rm CII}$\tablenotemark{a} & log N$_{\rm SiII}$\tablenotemark{a} & log N$_{\rm SiIII}$\tablenotemark{a} \\
     & ($\kms$)  & (cm$^{-2}$) & (cm$^{-2}$) & (cm$^{-2}$)  \\
\hline
S1 & -100, 100 & $>$15.13 & $>$14.98 & $>$14.11 \\
S2 &  -60, 140 & $>$15.17 & $>$15.02 & $>$14.02 \\
S3 & -100, 120 & $>$15.24 & $>$14.85 & $>$14.12 \\
S4 &  -80, 100 & $>$14.93 & $>$14.85 & $>$13.92 \\
S5 &  -80, 140 & $>$15.28 & $>$14.98 & $>$14.25 \\
S6 &  -80, 80  &      -\tablenotemark{b}   & $>$14.97 &      -    \\
S7 &  -80,  70 & $>$14.75 & $>$14.62 & $>$13.79 \\

\hline
\hline
\end{tabular}
\end{center}
\tablenotetext{a}{Since the absorption lines are all saturated, these column densities represent the lower limit. \ion{C}{2} column density is calculated using \ion{C}{2} $\lambda$1334, \ion{Si}{2} column density is calculated using \ion{Si}{2} $\lambda$1304 which is the weakest line with the least saturation among other \ion{Si}{2} lines, and \ion{Si}{3} column density is from \ion{Si}{3} $\lambda$1206. }
\tablenotetext{b}{See Appendix \ref{appC} for explanation of omitting \ion{C}{2} and \ion{Si}{3} along S6.}
\label{tb6}
\end{table}

%%%%%%%%% Table 6 end %%%% 

Here we show the properties of ions \ion{C}{2}, \ion{Si}{2}, and \ion{Si}{3} that have saturated absorption-line profiles. Fig \ref{figa1} shows the continuum-normalized line profiles. The portion of the absorption lines associated with M33's ISM is highlighted in gray shade based on visual inspection. The high degree of saturation limits our ability to discern the underlying individual kinematic structures, thus we do not attempt to decompose the M33 absorption lines. Instead, we apply the apparent optical depth (AOD) method (\citealt{Savage91, Savage96}) to calculate the column densities of \ion{C}{2}, \ion{Si}{2}, and \ion{Si}{3} that are associated with M33. 

The AOD method converts the observed absorption profiles into apparent column density as a function of velocity $N_{\rm a}(v)$,  which can be expressed as $N_{\rm a}(v)=3.768\times10^{14} \tau_{\rm a}(v) / (f \lambda({\rm \AA}))$, where $f$ and $\lambda$ are the oscillator strength and wavelength of the line of interest respectively, and $\tau_{\rm a}(v)$ is the apparent optical depth. For continuum-normalized absorption-line profiles, the apparent optical depth can be calculated as $\tau_{\rm a}(v)=-$ln[$I_{\rm obs}(v)/I_{\rm o(v)}$], where $I_{\rm o}(v)=1$ is the continuum intensity and $I_{\rm obs}(v)$ is the strength of the absorption lines. The total column density within a certain velocity range [$v_{\rm min}$, $v_{\rm max}$] is $N_{\rm a}(v)=\int_{v_{\rm min}}^{v_{\rm max}} N_{\rm a}(v) dv$. Since the \ion{C}{2}, \ion{Si}{2}, and \ion{Si}{3} absorption lines are strongly saturated, here the AOD method only provide lower limits of the ion column densities, which are shown in Table \ref{tb6}.

In addition to the saturated M33 and MW absorption lines, we find an $\vlsr\sim-350\kms$ ($-200\lesssim v_{\rm d} \lesssim -100\kms$) absorption feature that is commonly present in \ion{C}{2}, \ion{Si}{2}, and \ion{Si}{3} lines. Please see Section \ref{sec7.3.2} for the discussion of the potential origins of this feature.

\section{CalCOS V.S. Other Spectral Co-addition Pipelines}
\label{appB}

\begin{figure*}[b!]
\centering
\includegraphics[width=0.9\textwidth, height=0.8\textheight]{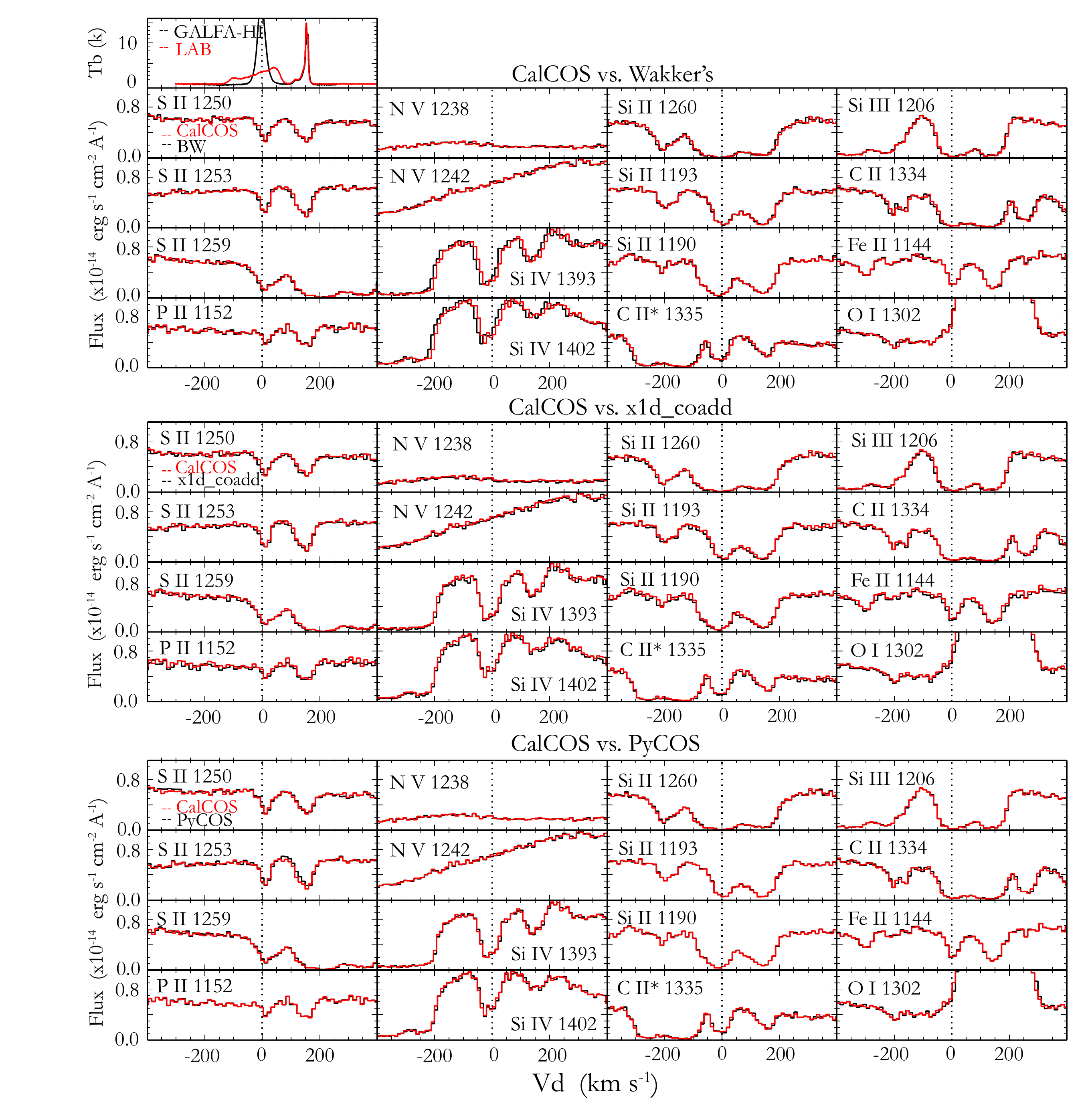}
\caption{Comparison of spectral alignment and co-addition results of S7 between CalCOS (red) and the other three pipelines (black) developed by \cite{Wakker15} (top), \cite{Danforth10} (middle) and C. Liang \& H.W. Chen (private communication; bottom), respectively. The panel on the top row shows the \ion{H}{1} 21-cm spectrum from the LAB (red; \citealt{Kalberla05}) and the GALFA-\ion{H}{1} (black; \citealt{Peek11}). The velocity on the x axis is with respective to the rotation velocity of the gas disk at the position of S7. Note that all the absorption lines are not continuum-normalized. }
\label{figb2}
\end{figure*}

As mentioned in Section \ref{sec2.2}, CalCOS pipeline yields an uncertainty in the wavelength solution of $14-19\kms$. Other spectral calibration and co-addition routines exist to minimize this wavelength uncertainty. To ensure that CalCOS products are reliable for our scientific analysis, we use three other pipelines to process the original raw {\it HST}/COS spectra and compare the results with that of CalCOS (version 3.0). The three pipelines are 1) x1d\_coadd.pro by \cite{Danforth10}, 2) a co-add code by \cite{Wakker15} and 3) the PyCOS pipeline by \cite{Liang14} (\& private communication). All the coadded spectra processed by these three methods can be found in \citet[][Dataset: \url{https://doi.org/10.5281/zenodo.168580}]{Zheng16}

First we use x1d\_coadd.pro \citep{Danforth10} to process the spectral co-adding for each of our seven sightlines. Since each sightline has four exposures, the program randomly chooses one exposure as a wavelength reference while the others are cross-correlated with it using strong ISM absorption lines over a 10 \AA\ spectral region. The code assumes a constant wavelength offset within the whole wavelength coverage. The final flux array is exposure-weighted average and the error array is exposure-weighted inverse invariance average. In Fig \ref{figb2}, we compare the co-added lines for S7 using this program with those from CalCOS. Good agreement is reached between the x1d\_coadd.pro-processed spectra and CalCOS products. Similar results are found for other sightlines. In the most extreme case, a $\sim$10$\kms$ offset is seen in S4.

B.P. Wakker kindly processed the raw spectra of S7 using his co-addition pipeline \citep{Wakker15} and provided us the final co-added product for comparison. His code is similar to x1d\_coadd.pro \citep{Danforth10} except that it cross-correlates all available strong ISM and IGM lines and applies relative offset for each individual line while \cite{Danforth10} assumes a constant shift for all the lines. The code aligns weakly-ionized ISM metal lines with the corresponding MW \ion{H}{1} 21 cm emission lines retrieved from the LAB survey \citep{Kalberla05}. We show the co-added spectra in Fig \ref{figb2}. Good alignment is found between CalCOS spectra and those by \cite{Wakker15} except that \ion{Si}{4} $\lambda\lambda$1393, 1402 show a $\sim10\kms$ displacement, which is less than one resolution element. In addition, the error array from his program is also consistent with that from CalCOS. 

Thirdly, we process the raw {\it HST}/COS data using the PyCOS pipeline developed by \cite{Liang14} (\& private communication). PyCOS provides a user-friendly interface with which one could process spectral alignment for individual lines interactively. The performance of line alignment is based on visual judgement and $\chi^2$ estimates. We use PyCOS to align and co-add the spectra, and also find good agreement between the PyCOS spectrum and the CalCOS one, as is shown in Fig \ref{figb2}. 

In all, our investigation shows a consistent result that the CalCOS pipeline provides reliable spectra for our scientific analysis. We have noted in Section \ref{sec2.3} that such good agreement between CalCOS and the other three pipelines is mainly due to the simple setup of our observations. For each of our sightlines, the observation was completed with four exposures in one single visit and the spectra were taken under the same setting. The background stars are all bright to ensure high S/N. Thus the possibility of spectral miss-alignment is largely reduced. CalCOS pipeline is most likely to become problematic in the case of faint QSO observations and that the spectra are taken at different epochs.

\section{Continuum fitting and Voigt-Profile Fitting}
\label{appC}

\begin{figure*}[t!]
\centering
\includegraphics[width=\textwidth]{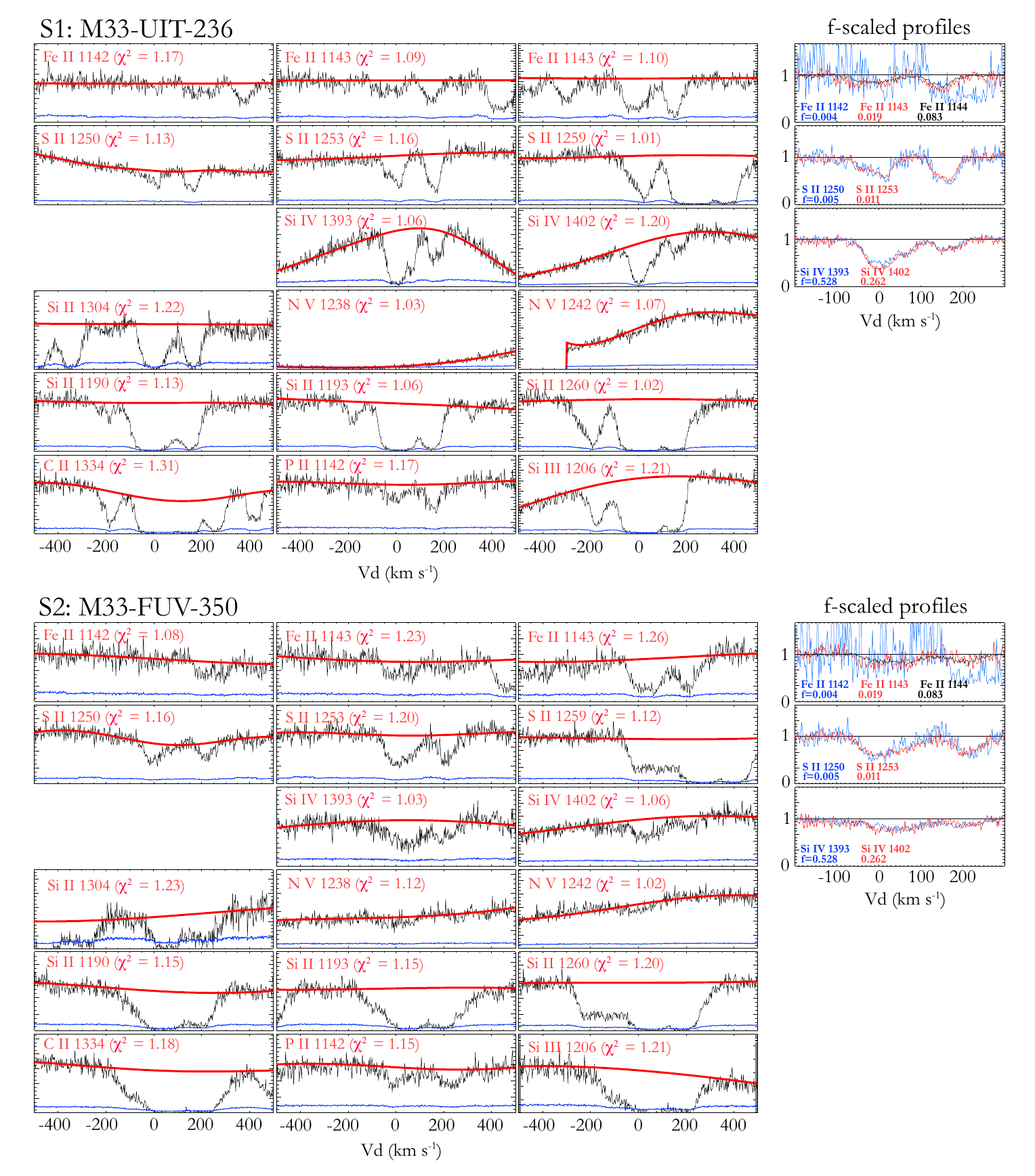}
\caption{Continuum fiting and $f$-scaled profiles of S1 (top) and S2 (bottom). The first three columns show the continuum fitting of each line, with black representing the original data, red for the fitted continuum and blue for the error. The right panels show the $f$-scaled profiles of \ion{Fe}{2}, \ion{S}{2}, and \ion{Si}{4}, respectively.}
\label{figc3}
\end{figure*}

\begin{figure*}[t!]
\centering
\includegraphics[width=\textwidth]{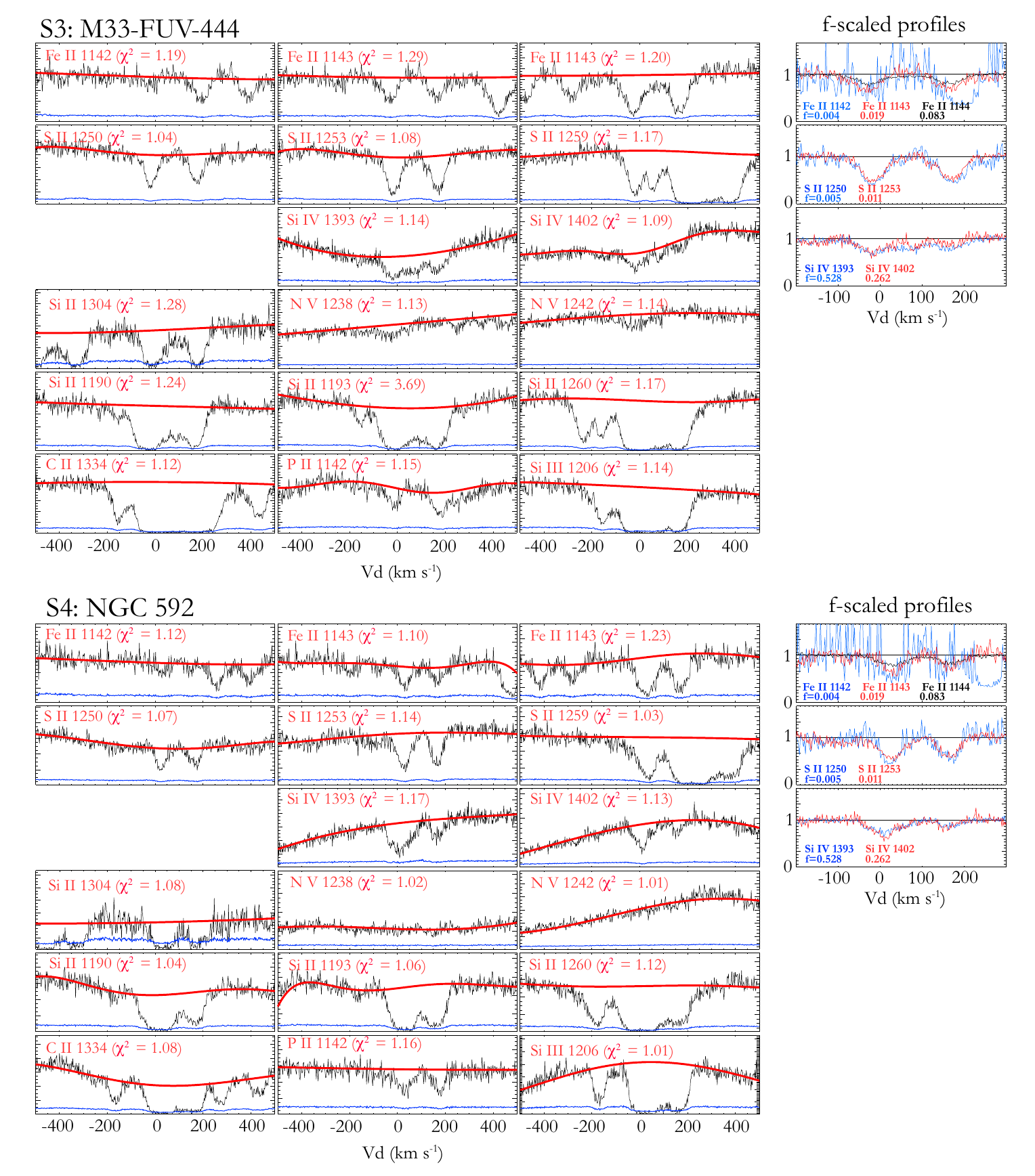}
\caption{Continuum fiting and $f$-scaled profiles of sightline S3 (top) and S4 (bottom). See Fig \ref{figc3} for layout explanation.}
\label{figc4}
\end{figure*}

\begin{figure*}[t!]
\centering
\includegraphics[width=\textwidth]{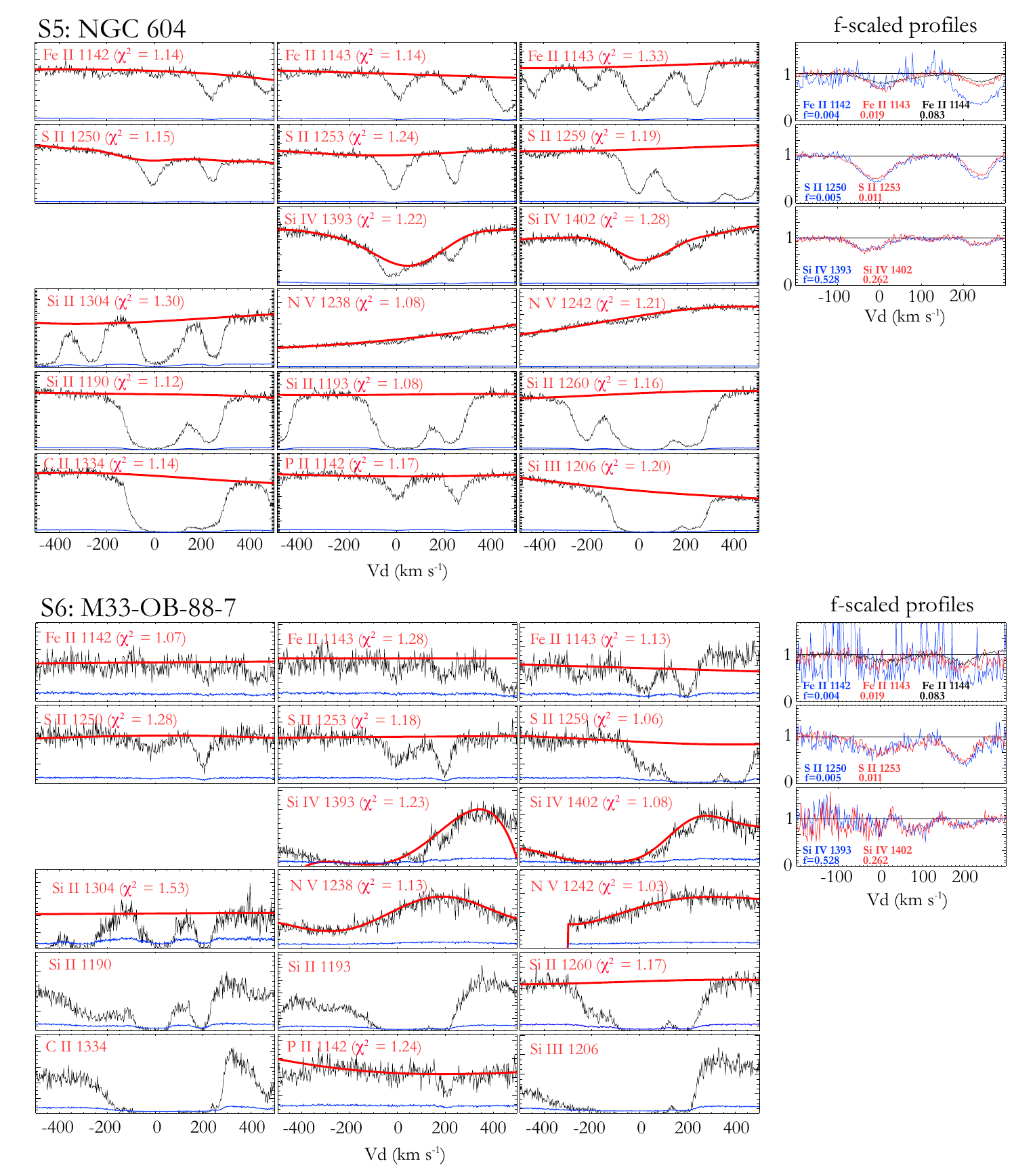}
\caption{Continuum fiting and $f$-scaled profiles of sightline S5 (top) and S6 (bottom). See Fig \ref{figc3} for layout explanation. In S6, \ion{C}{2} $\lambda$1334, \ion{Si}{2} $\lambda\lambda$1190, 1193, and \ion{Si}{3} $\lambda$1206 are not fitted since they are strongly affected by stellar P Cygni profiles. See Appendix \ref{appC} for detailed explanation. }
\label{figc5}
\end{figure*}

\begin{figure*}[t!]
\centering
\includegraphics[width=\textwidth]{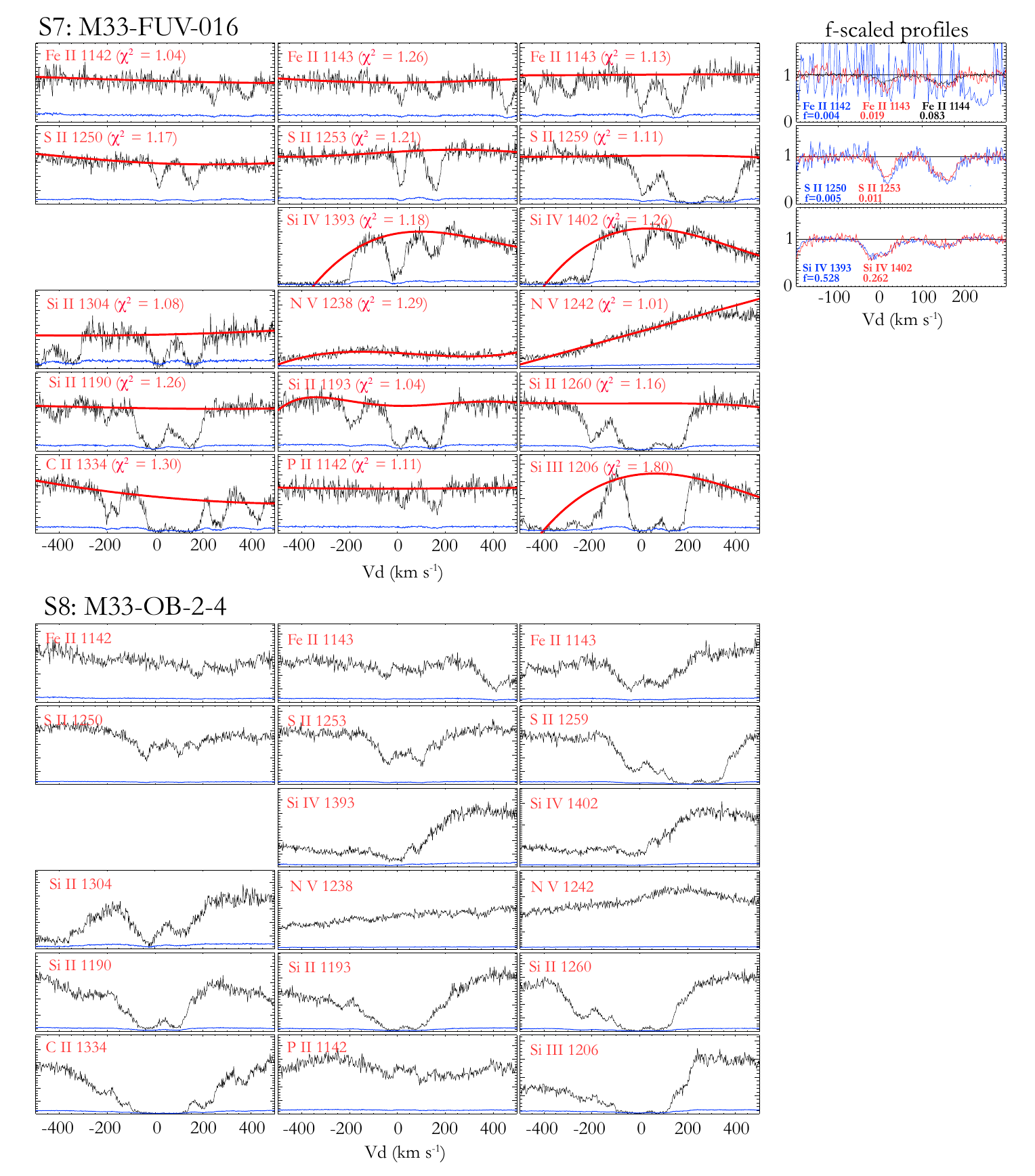}
\caption{Continuum fiting and $f$-scaled profiles of sightline S7 (top). See Fig \ref{figc3} for layout explanation. S8 is not used in our analysis. Here we show the original profile of each line to indicate the great uncertainity of the continuum fitting due to strong stellar activities. Please see Section \ref{sec2.1} for discussion of this sightline. }
\label{figc6}
\end{figure*}

Here we show the continuum fitting of each line for each target in Fig \ref{figc3}--\ref{figc6}. Most lines have good S/N, thus their continuum fitting and Voigt-profile fitting are straightforward. The $f$-scaled profiles in the right panels of Fig \ref{figc3}--\ref{figc6} show that the continuum fitting of most non-saturated lines are consistent, while some lines do show non-matching $f$-scaled profiles that we explain as follows: 

\begin{enumerate}

\item Along all the sightlines the $f$-scaled profiles of \ion{Fe}{2} $\lambda$1143 is stronger than it is supposed to be due to the blending with the \ion{Fe}{2} $\lambda$ 1142 from the MW. 

\item For S1, the $f$-scaled profiles of \ion{Si}{4} $\lambda\lambda$ 1393, 1402 lines do not match perfectly because \ion{Si}{4} $\lambda$ 1393 may be slightly saturated.

\item For the triplets \ion{S}{2} $\lambda\lambda\lambda$1250, 1253, 1259, we only use the first two lines for $f$-scaled matching since \ion{S}{2} $\lambda$1259 is severely blended with the $\vlsr\sim-350\kms$ feature from \ion{Si}{2} $\lambda$1260 as mentioned in Appendix \ref{appA} and \ref{sec7.3.1}. We do not include \ion{S}{2} $\lambda$1259 in our analysis. 

\item For sightline S6, we do not perform continuum fitting for \ion{C}{2} $\lambda$1334, \ion{Si}{2} $\lambda\lambda$1190, 1193 and \ion{Si}{3} $\lambda$ 1206. These lines are affected by strong P Cgyni profiles due to stellar activities. We show the original spectra of these lines in the bottom panel of Fig \ref{figc5}. The continuum peaks at $v_{\rm d}\sim300\kms$ in these lines are likely to be the emission components of the corresponding P Cygni profiles. These emission peaks can be clearly seen in panel 6 of Fig \ref{fig2}.  

\end{enumerate}

\begin{figure*}[t!]
\centering
\includegraphics[width=\textwidth]{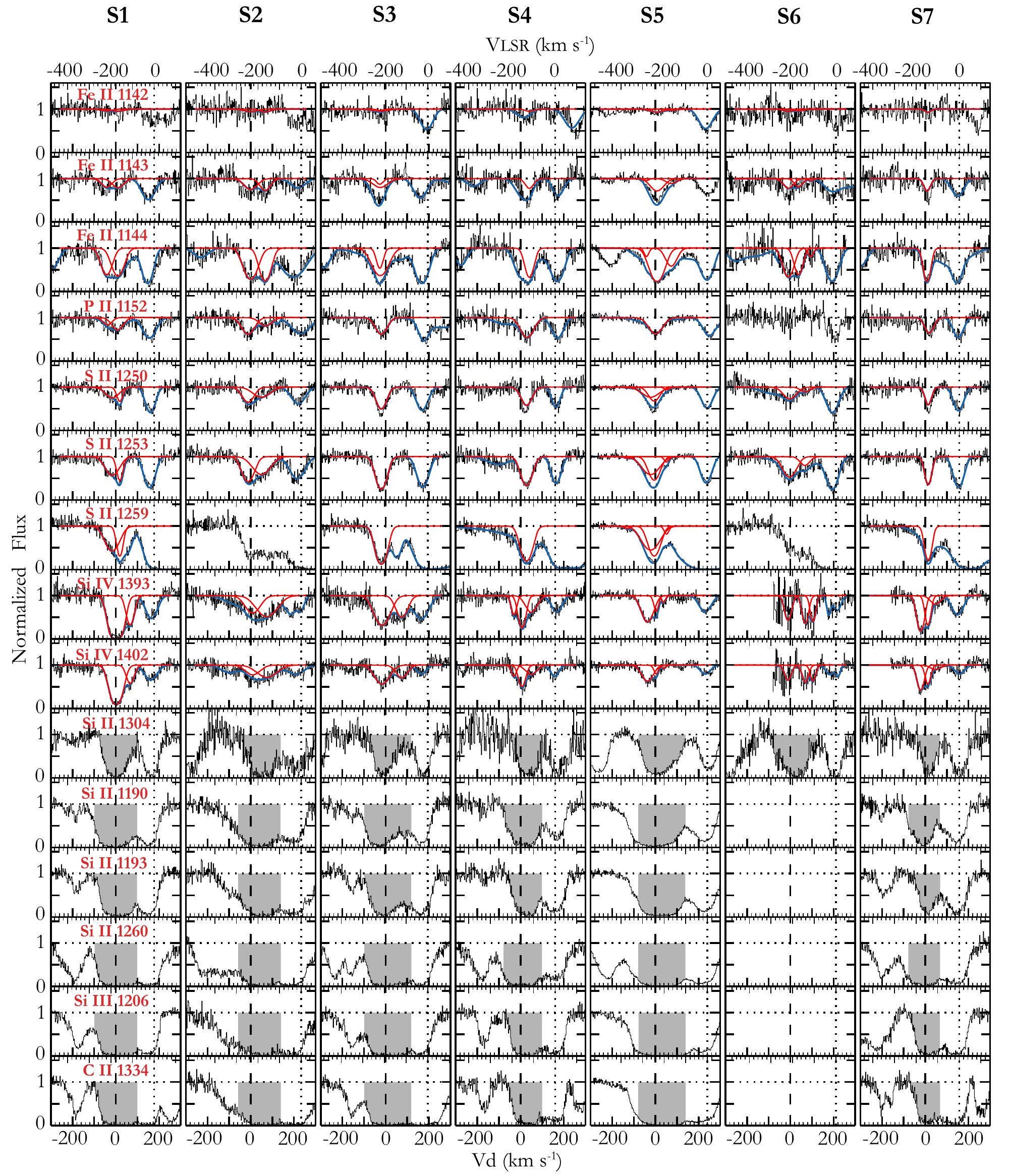}
\caption{Continuum-normalized spectra and Voigt-profile fitting for all the lines. M33's ISM absorption is shown by dashed line ($v_{\rm d}=0\kms$), and the MW's absorption is shown by dotted line ($\vlsr=0\kms$). The red curves indicate the Voigt-profile fits, and the blue curves show the overall fitting, including the sum of the red curves and the additional nuisance components (Appendix \ref{appC}). Gray shades indicate the integrated velocity range used in Appendix \ref{appA}.}
\label{figc7}
\end{figure*}

In Fig \ref{figc7}, we show all the non-saturated and saturated lines that were observed in our program. These includes those triplets and doublets that we do not show in Fig \ref{fig3} and \ref{figa1} for better illustration. In the following, we lay out the details of treatment of Voigt-profile fitting to some specific lines that we do not address in the previous sections:

\begin{enumerate}
\item On S3 and S4, the M33 \ion{Fe}{2} $\lambda$1143 is blended with the MW \ion{Fe}{2} $\lambda$1142, so a nuisance component is added to the Voigt-profile fitting. 

\item On S4, the M33 \ion{Fe}{2} $\lambda$1144 is blended with the MW \ion{Fe}{2} $\lambda$1143, so a nuisance component is added. 

\item On S6, the \ion{P}{2} $\lambda$1152 has no detection, thus we do not apply Voigt-profile fitting to this line. 

\item On all the sightlines \ion{S}{2} $\lambda$1259 is blended with a $\vlsr\sim-350\kms$ feature from \ion{Si}{2} $\lambda$1260, thus we do not include this line in our analysis. For some sightlines when the blending is less severe, we manage to fit Vogit-profile components to this line simultaneously with \ion{S}{2} $\lambda\lambda$1250, 1253. 
\end{enumerate}

\end{document}